\documentclass[bibyear]{aa}

\usepackage{silence}
\usepackage{graphicx}
\usepackage[singlelinecheck=false]{caption}

\usepackage[subrefformat=parens,labelformat=parens]{subcaption}
\usepackage{booktabs}
\usepackage{microtype}
\usepackage{placeins}

\usepackage[varg]{txfonts}

\usepackage{color}
\usepackage{natbib,twoopt}
\usepackage[breaklinks]{hyperref}
\bibpunct{(}{)}{;}{a}{}{,}
\definecolor{cobalt}{rgb}{0.06, 0.2, 0.65}
\hypersetup{
  colorlinks,
  citecolor=blue,
  linkcolor=blue,
  urlcolor=blue,
}
\makeatletter
  \newcommandtwoopt{\citeads}[3][][]{\href{http://adsabs.harvard.edu/abs/#3}%
    {\def\hyper@linkstart##1##2{}%
     \let\hyper@linkend\@empty\citealp[#1][#2]{#3}}}
  \newcommandtwoopt{\citepads}[3][][]{\href{http://adsabs.harvard.edu/abs/#3}%
    {\def\hyper@linkstart##1##2{}%
     \let\hyper@linkend\@empty\citep[#1][#2]{#3}}}
  \newcommandtwoopt{\citetads}[3][][]{\href{http://adsabs.harvard.edu/abs/#3}%
    {\def\hyper@linkstart##1##2{}%
     \let\hyper@linkend\@empty\citet[#1][#2]{#3}}}
  \newcommandtwoopt{\citeyearads}[3][][]%
    {\href{http://adsabs.harvard.edu/abs/#3}
    {\def\hyper@linkstart##1##2{}%
     \let\hyper@linkend\@empty\citeyear[#1][#2]{#3}}}
\makeatother

\usepackage{xcolor,soul}

\makeatletter
\renewcommand*\aa@pageof{, page \thepage{} of \pageref*{LastPage}}
\makeatother

\usepackage{pifont}
\usepackage{siunitx}
\DeclareSIUnit \magnitude {mag}
\DeclareSIUnit \mas {mas}
\DeclareSIUnit \Mjup {M\textsubscript{Jup}}
\DeclareSIUnit \Msun {M$_{\odot}$}
\DeclareSIUnit \parsec {pc}
\DeclareSIUnit \dex {dex}
\DeclareSIUnit \year {yr}
\DeclareSIUnit \au {au}

\newcommand{\cmark}{\ding{51}}%
\newcommand{\xmark}{\quad}%

\begin{document} 
\authorrunning{
    Hayoz et al.}
\titlerunning{
    ERIS}
\title{
    High-contrast spectroscopy with the new VLT/ERIS instrument}
\subtitle{
    Molecular maps and radial velocity of the gas giant AF~Lep~b\thanks{Based on observations collected at the European Organisation for Astronomical Research in the Southern Hemisphere under ESO program 112.2628.002.}}
\author{
J.~Hayoz\inst{\ref{ethz}}\fnmsep\thanks{\email{jhayoz@phys.ethz.ch}}
\and M.~J.~Bonse\inst{\ref{ethz}, \ref{mpiis}}
\and F.~Dannert\inst{\ref{ethz}, \ref{nccr}}
\and E.~O.~Garvin\inst{\ref{ethz}}
\and G.~Cugno\inst{\ref{uzh}}
\and P.~Patapis\inst{\ref{ethz}}
\and T.~D.~Gebhard\inst{\ref{ethz},\ref{mpiis}}
\and W.~O.~Balmer\inst{\ref{jhu}}
\and R.~J.~De~Rosa\inst{\ref{eso_santiago}}
\and A.~Agudo~Berbel\inst{\ref{mpe}}
\and Y.~Cao\inst{\ref{mpe}}
\and G.~Orban~de~Xivry\inst{\ref{star}}
\and T.~Stolker\inst{\ref{leiden}}
\and R.~Davies\inst{\ref{mpe}}
\and O.~Absil\inst{\ref{star}}\fnmsep\thanks{F.R.S.-FNRS Senior Research Associate}
\and H.~M.~Schmid\inst{\ref{ethz}}
\and S.~P.~Quanz\inst{\ref{ethz}, \ref{nccr}, \ref{eth_earth}}
\and G.~Agapito\inst{\ref{inaf_arcetri}}
\and A.~Baruffolo\inst{\ref{inaf_padova}}
\and M.~Black\inst{\ref{atc}}
\and M.~Bonaglia\inst{\ref{inaf_arcetri}}
\and R.~Briguglio\inst{\ref{inaf_arcetri}}
\and L.~Carbonaro\inst{\ref{inaf_arcetri}}
\and G.~Cresci\inst{\ref{inaf_arcetri}}
\and Y.~Dallilar\inst{\ref{koln}}
\and M.~Deysenroth\inst{\ref{mpe}}
\and I.~Di~Antonio\inst{\ref{inaf_abruzzo}}
\and A.~Di~Cianno\inst{\ref{inaf_abruzzo}}
\and G.~Di~Rico\inst{\ref{inaf_abruzzo}}
\and D.~Doelman\inst{\ref{leiden}}
\and M.~Dolci\inst{\ref{inaf_abruzzo}}
\and F.~Eisenhauer\inst{\ref{mpe}}
\and S.~Esposito\inst{\ref{inaf_arcetri}}
\and D.~Fantinel\inst{\ref{inaf_padova}}
\and D.~Ferruzzi\inst{\ref{inaf_arcetri}}
\and H.~Feuchtgruber\inst{\ref{mpe}}
\and N.~M.~F\"orster~Schreiber\inst{\ref{mpe}}
\and X.~Gao\inst{\ref{atc}}
\and R.~Genzel\inst{\ref{mpe}}
\and S.~Gillessen\inst{\ref{mpe}}
\and A.~M.~Glauser\inst{\ref{ethz}}
\and P.~Grani\inst{\ref{inaf_arcetri}}
\and M.~Hartl\inst{\ref{mpe}}
\and D.~Henry\inst{\ref{atc}}
\and H.~Huber\inst{\ref{mpe}}
\and C.~Keller\inst{\ref{leiden}}
\and M.~Kenworthy\inst{\ref{leiden}}
\and K.~Kravchenko\inst{\ref{mpe}}
\and J.~Lightfoot\inst{\ref{atc}}
\and D.~Lunney\inst{\ref{atc}}
\and D.~Lutz\inst{\ref{mpe}}
\and M.~MacIntosh\inst{\ref{atc}}
\and F.~Mannucci\inst{\ref{inaf_arcetri}}
\and T.~Ott\inst{\ref{mpe}}
\and D.~Pearson\inst{\ref{atc}}
\and A.~Puglisi\inst{\ref{inaf_arcetri}}
\and S.~Rabien\inst{\ref{mpe}}
\and C.~Rau\inst{\ref{mpe}}
\and A.~Riccardi\inst{\ref{inaf_arcetri}}
\and B.~Salasnich\inst{\ref{inaf_padova}}
\and T.~Shimizu\inst{\ref{mpe}}
\and F.~Snik\inst{\ref{leiden}}
\and E.~Sturm\inst{\ref{mpe}}
\and L.~Tacconi\inst{\ref{mpe}}
\and W.~Taylor\inst{\ref{atc}}
\and A.~Valentini\inst{\ref{inaf_abruzzo}}
\and C.~Waring\inst{\ref{atc}}
\and E.~Wiezorrek\inst{\ref{mpe}}
\and M.~Xompero\inst{\ref{inaf_arcetri}}
}

\institute{
ETH Zurich, Institute for Particle Physics and Astrophysics, Wolfgang-Pauli-Strasse 27, 8093 Zurich, Switzerland\label{ethz}
\and Max Planck Institute for Intelligent Systems, Max-Planck-Ring 4, 72076 Tübingen, Germany\label{mpiis}
\and National Center of Competence in Research PlanetS, Switzerland\label{nccr}
\and University of Zurich, Rämistrasse 71, 8006 Zurich, Switzerland\label{uzh}
\and Department of Physics \& Astronomy, Johns Hopkins University, 3400 N. Charles Street, Baltimore, MD 21218, USA\label{jhu}
\and European Southern Observatory, Alonso de C\'ordova 3107, Vitacura, Santiago, Chile\label{eso_santiago}
\and Max-Planck-Institut f\"ur extraterrestrische Physik, Postfach 1312, 85741, Garching, Germany\label{mpe}
\and Space Sciences, Technologies, and Astrophysics Research Institute, Universit\'e de Li\`ege, 4000 Sart Tilman, Belgium\label{star}
\and Leiden Observatory, Leiden University, Einsteinweg 55, 2333 CC Leiden, The Netherlands\label{leiden}
\and ETH Zurich, Department of Earth and Planetary Sciences, Sonneggstrasse 5, 8092 Zurich, Switzerland\label{eth_earth}
\and INAF -- Osservatorio Astrofisico di Arcetri, Largo E. Fermi 5, 50125, Firenze, Italy\label{inaf_arcetri}
\and INAF -- Osservatorio Astronomico di Padova, Vicolo dell'Osservatorio 5, 35122, Padova, Italy\label{inaf_padova}
\and STFC UK ATC, Royal Observatory Edinburgh, Blackford Hill. Edinburgh, EH9 3HJ, UK\label{atc}
\and I. Physikalisches Institut, Universit\"at zu K\"oln, Z\"ulpicher Str. 77, 50937, K\"oln, Germany\label{koln}
\and INAF -- Osservatorio Astronomico d'Abruzzo, Via Mentore Maggini, 64100, Teramo, Italy\label{inaf_abruzzo}
}
\date{Received 5 December 2024; accepted 20 March 2025}

\abstract
    {The Enhanced Resolution Imager and Spectrograph (ERIS) is the new adaptive optics (AO) assisted infrared instrument at the Very Large Telescope (VLT). Its refurbished integral field spectrograph (IFS) SPIFFIER leverages a new AO module, enabling high-contrast imaging applications and
    giving access to the orbital and atmospheric characterisation of super-Jovian exoplanets.}
    {We tested the detection limits of ERIS, and demonstrate its scientific potential by exploring the atmospheric composition of the young super-Jovian AF~Lep~b. Additionally, we improved its orbital solution by measuring its radial velocity relative to its host star.}
    {We present new spectroscopic observations of AF~Lep~b in $K$-band at $R\sim 11000$ obtained with ERIS/SPIFFIER at the VLT. We reduced the data using the standard pipeline together with a custom wavelength calibration routine, and removed the stellar point spread function using principal component analysis along the spectral axis. We computed molecular maps by cross-correlating the residuals with molecular spectral templates and measured the radial velocity of the planet relative to the star. Furthermore, we computed contrast grids for molecular mapping by injecting fake planets.}
    {We detect a strong signal from H$_{2}$O and CO but not from CH$_{4}$ or CO$_{2}$. This result corroborates the hypothesis of chemical disequilibrium in the atmosphere of AF~Lep~b. Our measurement of the RV of the planet yields $\Delta v_{\mathrm{R,P\star}} = \qty{7.8+-1.7}{\kilo\meter\per\second}$. This enables us to disentangle the degeneracy of the orbital solution; specifically, the correct longitude of the ascending node is $\Omega=248^{+0.4}_{-0.7}$~deg and the argument of periapsis is 
    $\omega=109^{+13}_{-21}$~deg.
    Our detection limits reach a contrast of $\Delta K = \qty{11.5}{\magnitude}$ at 0\farcs12 for the spectral templates of H$_{2}$O and CO, significantly extending the parameter space available to moderately high spectral resolution towards small angular separation.}
    {Our results demonstrate the competitiveness of the new ERIS/SPIFFIER instrument for the orbital and atmospheric characterisation of exoplanets at high contrast and small angular separation.}


\maketitle

\section{Introduction}
\label{section:introduction}

The combination of high-contrast imaging (HCI) and spectroscopy offers access to the orbital and atmospheric properties of low-mass companions and super-Jovian exoplanets.
Colloquially referred to as direct spectroscopy or high-contrast spectroscopy, this technique relies on the separation of the light coming from the companion from that of its host star using adaptive optics (AO) and post-processing algorithms that disentangle their significantly different spectral signatures \citep{Snellen2015CombiningNeighbors}. 

Early results of direct spectroscopy include the detection of carbon monoxide in the atmosphere of the giant exoplanet HR~8799~c \citep{Konopacky2013DetectionAtmosphere} using the OH-Suppressing Infrared Integral Field Spectrograph \citep[OSIRIS;][]{Quirrenbach2003Integral-fieldKeck,Larkin2006OSIRIS:Keck} instrument at the Keck telescope as well as the measurement of the spin velocity of $\beta$~Pic~b \citep{Snellen2014FastPictorisb} using the Cryogenic High-Resolution Infrared Echelle Spectrograph \citep[CRIRES;][]{Kaeufl2004CRIRESVLT} from the Very Large Telescope (VLT). Since then, many studies have continued to develop this technique further: OSIRIS has enabled the measurement of the radial velocity (RV) of the HR~8799~b and c planets \citep{Ruffio2019RadialSpectroscopy} and the constraint of the atmospheric parameters of the c planet \citep{Wang2023RetrievingData}. The Spectrograph for INtegral Field Observations in the Near Infrared \citep[SINFONI;][]{Eisenhauer2003SINFONIVLT} at the VLT was used to measure the atmospheric $^{12}$CO/$^{13}$CO isotopologue ratio of the young giant exoplanet TYC~8998-760-1~b \citep{Zhang2021TheSuper-Jupiter}. Using the NIRSPEC spectrograph \citep{McLean1998DesignTelescope,Lopez2020CharacterizationTelescope,Fitzgerald2018AnTelescope} from the Keck Planet Imager and Characterizer \citep[KPIC;][]{Delorme2021KeckSpectroscopy}, \citet{Xuan2022ASpectroscopy} measured the carbon-to-oxygen ratio (C/O) and metallicity of the brown dwarf HD~4747~B. More recently, \citet{Ruffio2024JWST-TSTUnit} measured the \qtyrange{2.9}{5.2}{\micro\meter} spectrum of the T~dwarf HD~19467~B using the NIRSpec \citep{Jakobsen2022TheTelescope,Boker2023In-orbitTelescope} integral field spectrometer (IFS) on board the James Webb Space Telescope (JWST), leading to the detection of H$_{2}$O, CH$_{4}$,  CO$_{2}$, and CO in its atmosphere \citep{Hoch2024JWST-TSTSpectrograph}. Beyond the exploration of the atmospheric physics of gas giant exoplanets, the atmospheric composition of low-mass companions might provide insights into their formation history \citep[see, e.g.,][]{Oberg2011THEATMOSPHERES,Oberg2021AstrochemistrySystems,Turrini2021TracingSulfur,Pacetti2022ChemicalPlanets}. While the details of planet formation are still mostly unclear, these measurements, which are enabled by newer instruments and more powerful analysis tools, might help to put together some pieces of the puzzle.
\looseness=-1

The latest facility offering high-contrast spectroscopic capabilities on an \qty{8}{\meter} class telescope is the Enhanced Resolution Imager and Spectrograph \citep[ERIS;][]{Davies2023TheVLT}. Installed at the Cassegrain focus of the Unit Telescope 4 (UT4) at the VLT, ERIS is the new AO-assisted instrument replacing the Infrared imaging and spectroscopic capabilities previously offered by NAOS-CONICA \citep[NACO;][]{Lenzen2003NAOS-CONICAModes,Rousset2003NAOSPerformance} and SINFONI \citep{Eisenhauer2003SINFONIVLT}. With its refurbished and upgraded IFS SPIFFIER \citep{George2016MakingRe-commissioning} as well as its new AO module \citep{Riccardi2022TheVLT-UT4}, ERIS is expected to improve upon its predecessor SINFONI in terms of contrast, angular separation, and sensitivity. However, its competitiveness in the current landscape of high-contrast spectrographs remains to be proven.
In this article, we aim to demonstrate the scientific potential of this new instrument by reporting on the detection of one of the most challenging directly imaged exoplanets discovered so far, AF~Lep~b, which sits at an angular separation of 0\farcs32 and contrast of $\Delta K = \qty{11.84}{\magnitude}$ \citep{DeRosa2023DirectLeporis}.
\looseness=-1

The super-Jovian exoplanet AF~Lep~b was simultaneously discovered by three teams who were independently following up on the significant astrometric accelerations of its host star measured between the \textit{Gaia} and \textit{Hipparcos} catalogues \citep{DeRosa2023DirectLeporis,Mesa2023AFData,Franson2023AstrometricLep}. The exoplanet was later precovered in archival NACO observations taken 11 years prior \citep{Bonse2025Use2011}, and since then has been re-observed with the JWST Near Infrared Camera \citep[NIRCam;][]{Franson2024JWST/NIRCamB} and GRAVITY at the VLT Interferometer \citep[VLTI;][]{Balmer2025VLTI/GRAVITYMetallicity}. Its dynamical mass was measured at \qty{3.75+-0.5}{\Mjup} \citep{Balmer2025VLTI/GRAVITYMetallicity}, making it one of the lowest-mass directly imaged exoplanets found to date. By combining the relative astrometry measured by GRAVITY with all archival RV and astrometric data, most of the orbital parameters could be accurately determined: its period is $24.3^{+0.9}_{-0.4}$~yr, its semi-major axis is $8.98^{+0.15}_{-0.08}$~au and its eccentricity is smaller than 0.02 \citep{Balmer2025VLTI/GRAVITYMetallicity}.
AF~Lep is an F8V star \citep{Gray2006ContributionsSample} at \qty{26.8}{\parsec} with a mass of $1.13^{+0.10}_{-0.04}$~M$_{\odot}$ \citep{DeRosa2023DirectLeporis}. It is a likely member of the $\beta$~Pictoris moving group \citep{Zuckerman2001TheGroup,Malo2013BAYESIANGROUPS,Ujjwal2020AnalysisDR2} with an age of \qty{24+-3}{\mega\year} \citep{Bell2015ANeighbourhood}. 
The latest atmospheric study---based on the currently available photometry and low-resolution spectroscopy---locate the planet at the L--T transition and indicate an effective temperature of $T_{\mathrm{eff}} = \qty{800+-50}{\kelvin}$ with evidence of elevated metallicity and chemical disequilibrium \citep{Balmer2025VLTI/GRAVITYMetallicity}, which agree with previous atmospheric constraints \citep{Zhang2023ELementalAtmospheres,Palma-Bifani2024AtmosphericModeling,Franson2024JWST/NIRCamB}.
\looseness=-1

The combined fit on the GRAVITY relative astrometry together with RV and astrometric data was not able to constrain two of the orbital parameters AF~Lep~b \citep{Balmer2025VLTI/GRAVITYMetallicity}.
Since direct imaging, and hence interferometry, projects the astrometric path of astronomical objects on the celestial sphere, it is insensitive to the component of the velocity of the planet that is orthogonal to the sky plane, and therefore results in a bimodal distribution for the ascending node and argument of periapsis. This ambiguity prevents the prediction of the orbital phase and the determination of the true obliquity of the planet. The former is relevant for planning follow-up observations in reflected light, while the latter can inform the planet formation history of AF~Lep~b \citep{Kraus2020SpinOrbitSystem}. Orbital motion perpendicular to the sky plane can only be determined by RV measurements. Due to the fast rotation of the star, existing stellar RV measurements of AF~Lep are not accurate enough to robustly resolve the reflex motion of the star caused by the planet \citep{DeRosa2023DirectLeporis}. Therefore, the only way to uniquely disambiguate the orbital solution is by measuring the RV of the planet directly.
\looseness=-1

In this article we present new ERIS/SPIFFIER $K$-band observations of AF~Lep~b at moderately high spectral resolution ($R \sim \num{11000}$) obtained within the ERIS Guaranteed Time Observations (GTO) programme allocated to ETH Zurich. We describe the observations and our data reduction (Sect.~\ref{section:methods:observations} and Sect.~\ref{section:methods:data_reduction}), including our custom wavelength calibration (Sect.~\ref{section:methods:wavelength}) as well as our algorithm to remove the stellar point spread function (PSF; Sect.~\ref{section:methods:psf_subtraction}). We compute molecular maps \citep{Hoeijmakers2018Medium-resolutionImaging} of the AF~Lep system (Sect.~\ref{section:results:molmaps}), revealing the planet through the detection of water and carbon monoxide in its atmosphere. We then describe how we measured the RV of AF~Lep~b relative to its star (Sect.~\ref{section:results:RV}), which allowed us to resolve the ambiguity in its orbital solution (Sect.~\ref{section:discussion:orbital_constraint}). We discuss the implications of our detection of H$_{2}$O and CO (Sect.~\ref{section:discussion:chemistry}), the impact of stellar contamination (Sect.~\ref{section:discussion:stellar_contamination}), and compute the detection limits for our observations (Sect.~\ref{section:discussion:contrast_performance}). In Sect.~\ref{section:conclusion} we provide our concluding remarks.
\looseness=-1

\section{Methods}
\label{section:methods}

\subsection{Observations}
\label{section:methods:observations}

\begin{figure*}[ht]
\centering
    \includegraphics[height=5cm]{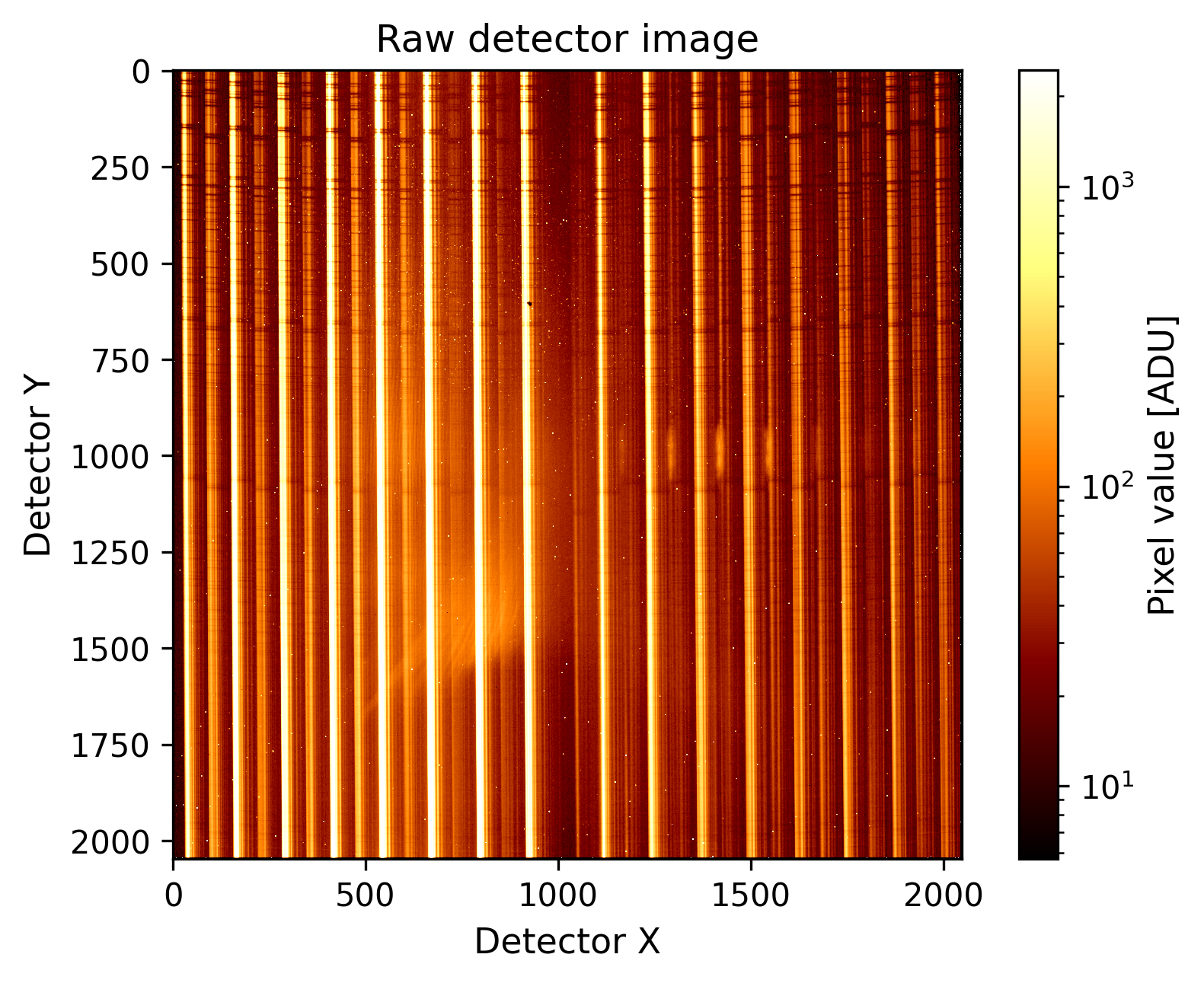}
    \includegraphics[height=5cm]{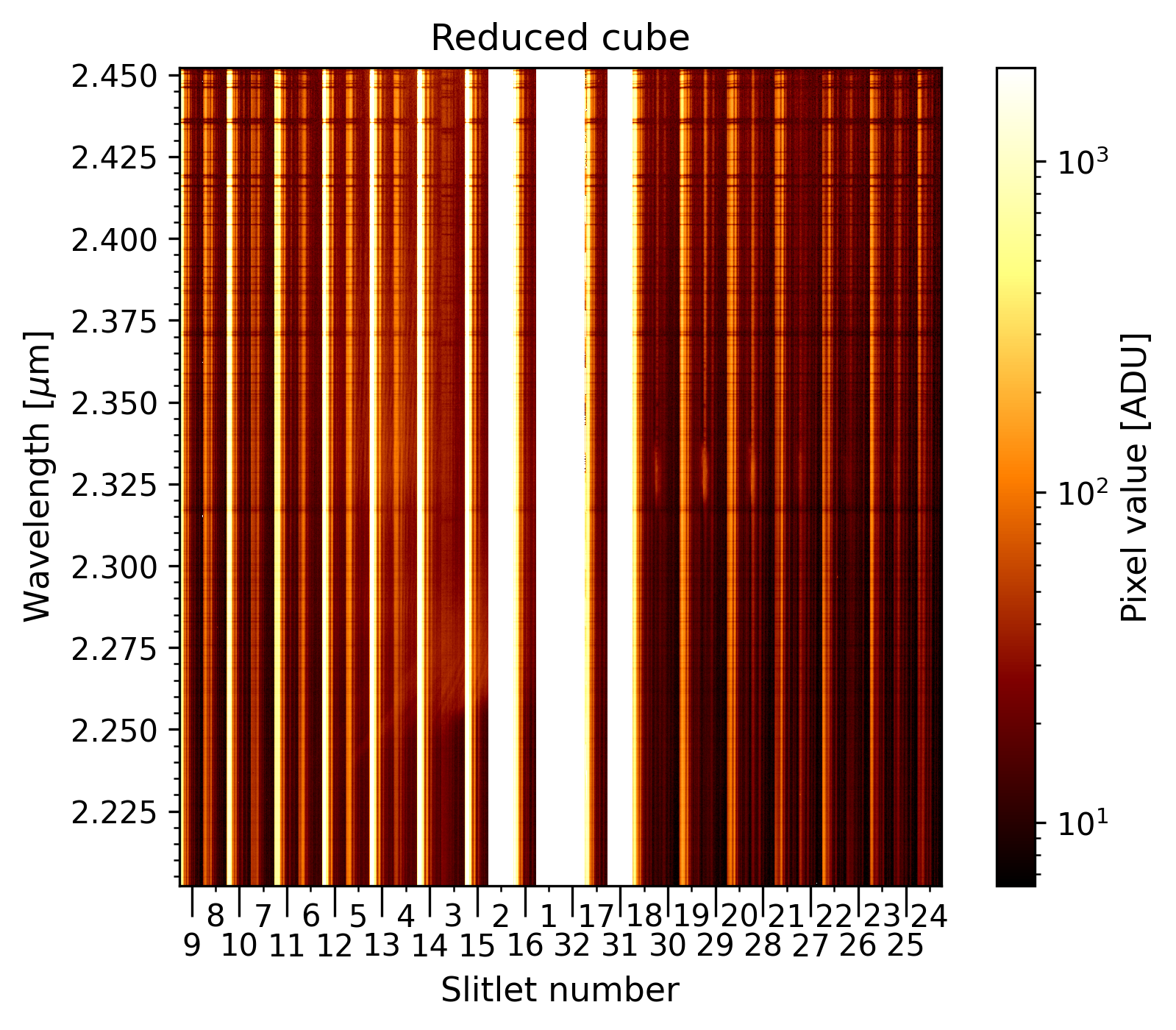}
    \includegraphics[height=5cm]{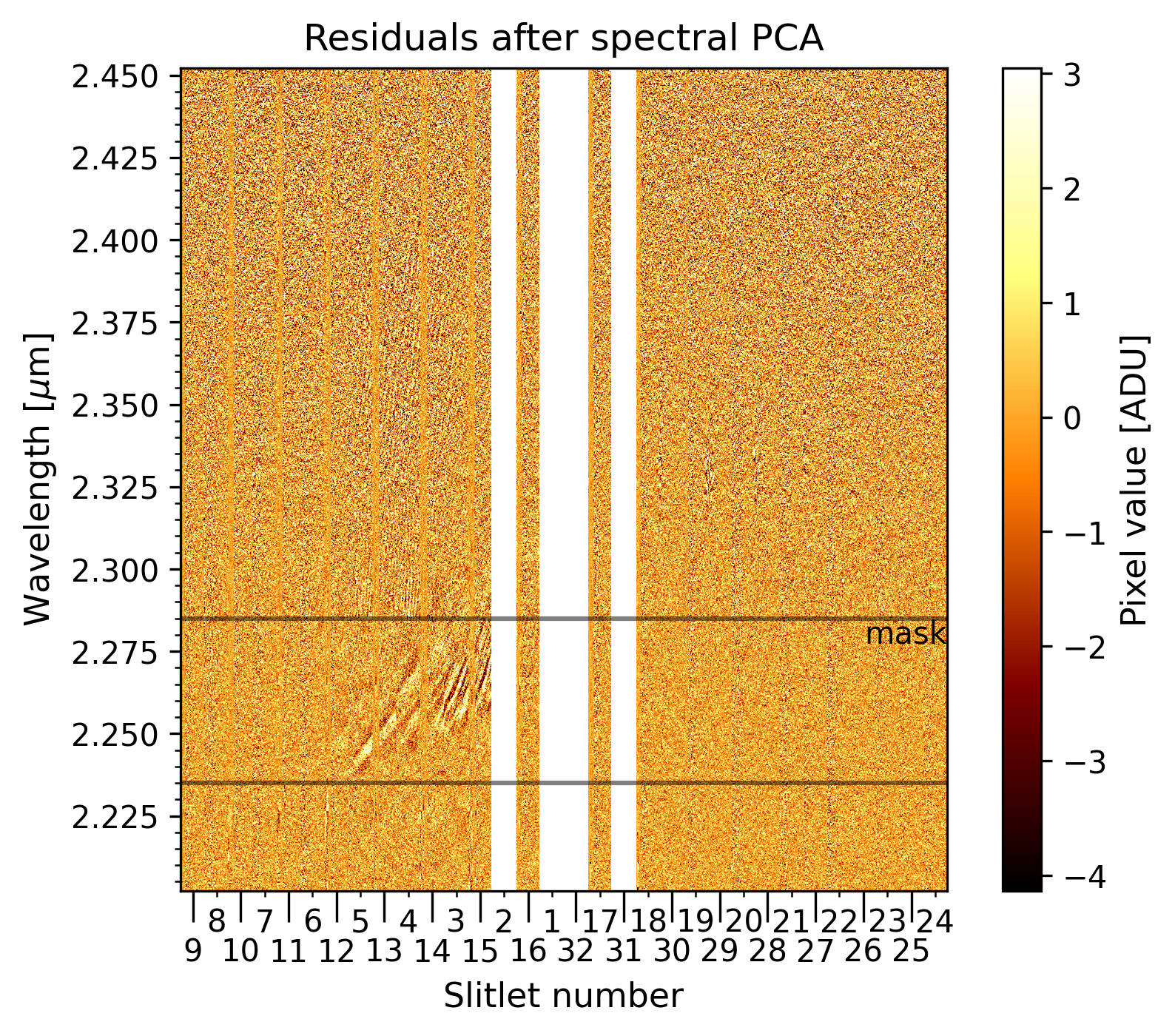}
\caption{Illustration of the main processing steps applied to one frame of our dataset. \textit{Left:} Raw detector frame. The brick-wall pattern of the slitlets is visible as the telluric absorption lines are shifted in adjacent slitlets. \textit{Middle:} Reduced frame. The missing data in the middle is due to cropping the frames, essentially removing slitlets at the border of the image. \textit{Right:} Residuals after PSF subtraction via spectral PCA. The two horizontal lines at \num{2.24} and \qty{2.30}{\micro\meter} mark the position of the mask applied to the data to hide the strong stray light features present in slitlets 2 to 6 and 12 to 15 (cf. Fig.~\ref{fig:spectrum}). The slitlets are numbered from 1 to 32, and span the image from bottom to top (i.e. from south to north).}
\label{fig:illustration_data}
\end{figure*}

AF~Lep was observed with ERIS on the  8  November 2023 using SPIFFIER with the smallest image scale ($0\farcs8 \times 0\farcs8$) and the K-long grating (\qtyrange{2.19}{2.47}{\micro\meter} at $R \sim \num{11000}$). The observations were taken in field-tracking mode (pupil-tracking is not  available at this time) with position angle set to \qty{0}{\degree} (i.e.  north at the top of the image); the star was offset by 0\farcs15 to the west to include the exoplanet in the field of view, with a square jitter pattern of width $0\farcs1$ with offsets every second frame, and sky offsets after each square jitter. We recorded a total of \num{267} object frames with a detector integration time (DIT) of \qty{40}{\second} and $n_{\mathrm{DIT}}=1$. The observing conditions were average, with a median seeing of $0\farcs65$, a median coherence time of \qty{5.7}{\milli\second}, and a median integrated water vapour column of \qty{3.32}{\milli\meter}. 

\subsection{Data reduction}
\label{section:methods:data_reduction}

The data were reduced using the ERIS-SPIFFIER pipeline v1.6.0 together with custom tools to improve the quality of the calibration. The standard calibration steps consist of dark subtraction,  flat fielding, bad pixel correction, distortion correction, and an initial wavelength solution using daytime arc lamp calibrations. The pipeline normally relies on the OH sky emission lines present in the science data to obtain a good wavelength calibration \citep{Davies2007ASpectra}; however, our observations are dominated by the stellar PSF, and the OH lines are not visible in the data, which prevented us from using this calibration strategy. Therefore, we had to implement a custom tool to calibrate the wavelength using the telluric absorption lines visible in the stellar PSF. We give a short summary of our custom wavelength calibration in Sect.~\ref{section:methods:wavelength}, while the full description can be found in Appendix~\ref{app:wavelength_calibration}. We describe the additional pre-processing steps necessary to format the output of the ERIS-SPIFFIER pipeline in Sect.~\ref{section:methods:additional}, and our PSF subtraction technique in Sect.~\ref{section:methods:psf_subtraction}. The main steps of our pipeline---including the PSF subtraction---are illustrated in Fig.~\ref{fig:illustration_data}. In the following, we use the term spaxel to denote the spectrum extracted from one cube at a given spatial pixel.
\looseness=-1

\subsubsection{Custom wavelength calibration}
\label{section:methods:wavelength}

The ERIS instrument is installed in the Cassegrain focus of the UT4 telescope. Consequentially, internal flexures inevitably impact the position of the spectrum on the detector as the instrument rotates and changes altitude to track its target. This renders daytime wavelength calibrations inaccurate and necessitates   leveraging the science frames to produce an accurate wavelength solution. For observations of faint targets with long integration times and large field of views, the ERIS-SPIFFIER pipeline makes use of the OH sky emission lines as references to compute an accurate wavelength solution \citep{Davies2007ASpectra} for each science frame. As mentioned above, this is not possible in our case due to the stellar PSF dominating the full frame and hiding the OH emission lines. Instead, we can leverage the telluric absorption lines imprinted on the stellar spectrum to improve the wavelength calibration. As reference, we used a telluric transmission template calculated with \texttt{SkyCalc} \citep{Noll2012AnParanal,Jones2013AnParanal} and evaluated it with the parameters matching our observing conditions. Our custom wavelength calibration consists of three steps, which are summarised below.

Step 1: We first measure an initial shift in each spaxel of the data by cross-correlating each spaxel with the telluric transmission template. This first step accounts for the spectral curvature, which causes a line of constant wavelength to follow a curve across the detector. 

Step 2: We measure higher-order wavelength errors. To do this, we first correct the spectral curvature measured in the first step and combine all spaxels within the same slitlet. We then measure the local error by cross-correlating the data and the telluric transmission template on a moving window along the wavelength axis. This measurement yields the wavelength error as a function of wavelength, which we fit using a spline.

Step 3: We update the wavelength solution. This is done by adding the wavelength error measured in the previous steps to the initial wavelength map computed by the ERIS-SPIFFIER pipeline from arc lamp images taken during daytime calibrations. We then run the \texttt{eris\_ifu\_jitter} recipe together with the updated wavelength map, which produces the calibrated datacubes.

\begin{figure}[t]
    \centering
    \includegraphics[width=\hsize]{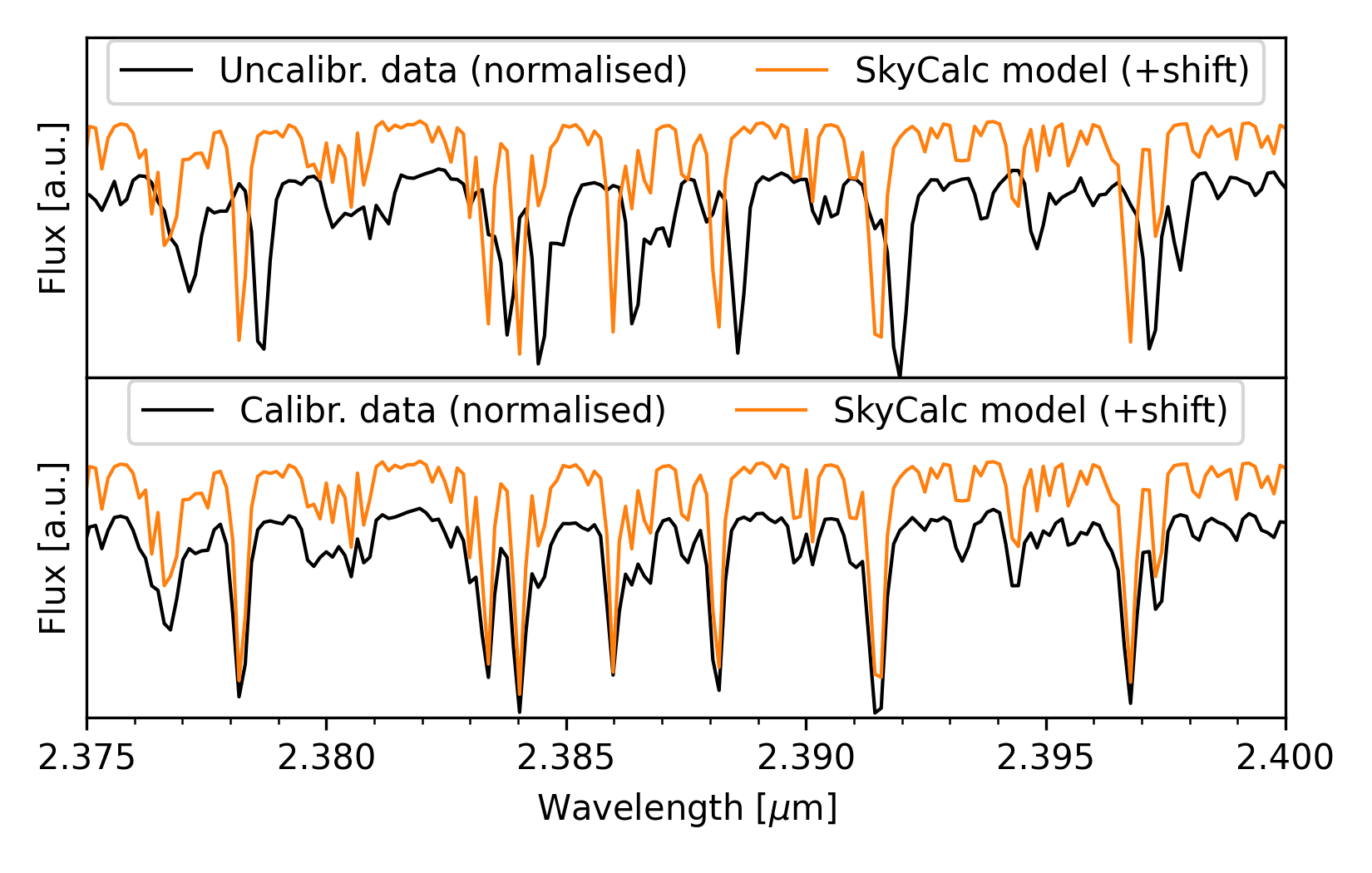}
    \caption{Application of our custom wavelength calibration to the spectrum.
    \textit{Upper panel:} Spectrum calibrated using daytime arc lamp images as provided by the standard pipeline when OH sky emission lines are hidden by the stellar PSF.
    \textit{Lower panel:} Spectrum calibrated using our custom pipeline. We provide the telluric transmission template used for the calibration as reference (orange). We only show the region between \qtyrange{2.375}{2.4}{\micro\meter} to easily distinguish the telluric lines by eye.}
    \label{fig:wavelength_calibrated_spectrum}
\end{figure}

Figure~\ref{fig:wavelength_calibrated_spectrum} shows a small region of the mean spectrum contained within the slitlet 11 before and after our wavelength calibration together with the telluric transmission template for comparison. The improvement is clearly visible by eye: the telluric lines present in the calibrated data match the template much more closely. In Appendix~\ref{app:wavelength_calibration}, we give a more exhaustive description of our wavelength calibration procedure (see Fig.\ref{fig:wavelength_calibration} for an illustration of the two main steps) and estimate its accuracy.

\subsubsection{Pre-processing}
\label{section:methods:additional}

After the wavelength calibration, each exposure is a 3D cube consisting of \num{2155} $\lambda$-images of size $64\times64$, covering a wavelength range between \qtyrange[range-phrase=~and~]{2.18799}{2.46814}{\micro\meter}. We implemented a custom pipeline to prepare the data for post-processing. Due to the brick-wall arrangement of the slitlets on the detector (see the left image in Fig.~\ref{fig:illustration_data}), we excluded all $\lambda$-images outside the range \qtyrange{2.20229}{2.45176}{\micro\meter} to avoid any missing data. Furthermore, we observed the presence of bright stray light extending between \qtyrange{2.24}{2.30}{\micro\meter} and within slitlets 2 to 6 and 12 to 15 (see Fig.~\ref{fig:illustration_data}). The cause for this stray light within ERIS/SPIFFIER is unclear. Since it effectively adds strong correlated noise along the spectral and spatial dimensions, it can hinder the detection of a molecular signal in the atmosphere of a companion or create fiducial features that might be mistaken for one. Therefore, we masked the full spectral region between \qtyrange{2.24}{2.30}{\micro\meter}. We further removed \num{4} pixels at the edge of each frame to avoid contamination by adjacent slitlets due to inaccuracies of the distortion calibration, leaving $\lambda$-images of size $56\times56$. Due to residual instrument flexures between the AO stage and the SPIFFIER detector at the moment of our observations, the target was slowly drifting over time at a rate of approximately \qty{60}{\mas\per\hour}. This drift forced us to pause the observation after \qty{1.5}{\hour} to re-centre the star. Nevertheless, we still had to remove \num{28} frames for which the star drifted outside the field of view, leaving \num{239} frames. The position of the star in the remaining frames was measured by fitting the stellar PSF with a Gaussian in the wavelength-averaged frames. At this point, the data had reached their final shape; there were \num{239} cubes left, each made of \num{1923} $\lambda$-images of size $56\times56$.\looseness=-1

After PSF subtraction but before cross-correlating the data with spectral templates (cf. sections~\ref{section:methods:psf_subtraction} and \ref{section:results:molmaps}), we convolved each image with an aperture of radius \qty{1.7}{px} (\qty{21.25}{\mas}), i.e., we applied aperture photometry on every pixel. The output of this step is a datacube with the same format as the input cube, but where each spaxel now contains the total flux over its surrounding aperture. This step was carried out to gather all the signal coming from the planet and thereby increasing the signal-to-noise ratio (S/N) of its spectrum. We tried several radii for the apertures, including 1\,$\lambda/D$ (i.e. \qty{58.5}{\mas}); however, they yielded worse results, probably because of remaining systematics from the PSF subtraction that are included in the apertures. The spectrum of the planet is extracted after this convolution by taking the spaxel resulting in the highest S/N in the molecular maps.

\begin{table*}[ht]
    \centering
    \renewcommand*{\arraystretch}{1.2}
    \caption[]{Removal of systematic noise with PCA.}
    \label{tab:pca_methods}
     \begin{tabular}{l c c c c c c l} 
        \hline
        Algorithm & Data type & $R$ & Usage of PCA & Features & Samples & Forward Model & References \\
        \hline
        ADI + PCA/KLIP & Imaging & -- & Full PSF & $X \times Y$ & $T$ & \xmark & (1), (2) \\
        TRAP & Imaging & -- & Full PSF & $T$ & $X \times Y$ & \cmark & (3) \\
        PCA-Temporal & Imaging & -- & Full PSF & $T$ & $X \times Y$ & \xmark & (4) \\
        ASDI/CODI & IFS & $< \num{1000}$ & Full PSF & $X \times Y$ & $\Lambda \times T$ & \xmark & (5) \\
        FMMF & IFS & $< \num{1000}$ & Full PSF & $X \times Y$ & $\Lambda \times T$ & \cmark & (6) \\
        HRSDI & IFS & $> \num{1000}$ & Residual systematics & $\Lambda$ & $X \times Y$ & \xmark & (7), (8), (9) \\
        Ruffio and Hoch & IFS & $> \num{1000}$ & Residual systematics & $\mathcal{P} \times \Lambda$ & $X \times Y$ & \cmark & (10), (11) \\
        \hline
        Spectral PCA & IFS & $> \num{1000}$ & Full PSF & $\Lambda$ & $X \times Y$ & \xmark & This work \\
        \hline
     \end{tabular}
     \tablefoot{Overview of application schemes for PCA-based systematic noise subtraction. Depending on the scheme and data type, PCA can be applied along different dimensions: 
     the letters $X$ and $Y$ denote the two spatial axes, $T$ is the time axis, $\Lambda$ is the spectral axis, and $\mathcal{P}$ is a local $5 \times 5$ patch in space. The column $R$ denotes the spectral resolution. In the last three rows, i.e. for the methods HRSDI, Ruffio and Hoch, and Spectral PCA, the column "Usage of PCA" refers to the steps (i) and (ii) described in the main text.
     References: 
     (1) \citet{Amara2012scppynpoint/scpExoplanets}, 
     (2) \citet{Soummer2012DETECTIONEIGENIMAGES}, 
     (3) \citet{Samland2021TRAP:Separations}, 
     (4) \citet{Long2023ImprovedTechniques}, 
     (5) \citet{Kiefer2021SpectralSPHERE/IFS}, 
     (6) \citet{Ruffio2017ImprovingFilter},
     (7) \citet{Hoeijmakers2018Medium-resolutionImaging}, 
     (8) \citet{Haffert2019Two70},
     (9) \citet{Cugno2021MolecularSystem},
     (10) \citet{Hoch2020Moderate-resolutionB},
     (11) \citet{Ruffio2021DeepSpectroscopy}.}
\end{table*}

\subsubsection{Removal of the stellar point spread function with spectral PCA}
\label{section:methods:psf_subtraction}

Since the stellar PSF dominates the signal from the exoplanet, we needed to model and subtract it from the data. As pupil-tracking was not available at the time of our observations, we could not use angular differential imaging \citep[ADI;][]{Marois2006AngularTechnique}. 
Instead, we wanted to leverage the significant differences in spectral features (which can be resolved at high enough spectral resolution) between the companion and the tellurics-imprinted stellar spectrum. This was able to be done by using principal component analysis \citep[PCA; see e.g.][]{bishop2006pattern}.

PCA is widely used by the exoplanet community to suppress systematic noise. Depending on the type of data, different schemes have been developed to apply PCA.
In general, PCA models the strongest covariances along what is known as the feature dimension using a set of data points called samples.
In the context of HCI, PCA was first introduced to ADI \citep[][]{Amara2012scppynpoint/scpExoplanets,Soummer2012DETECTIONEIGENIMAGES}. 
In this framework, PCA explores the spatial covariances of the speckle noise utilising the image dimensions ($X \times Y$) as features and the time dimension ($T$) as samples. 
An alternative scheme to apply PCA is \texttt{PCA-Temporal} \citep{Long2023ImprovedTechniques} or similarly \texttt{TRAP} \citep{Samland2021TRAP:Separations}.
Instead of exploring the spatial covariances, PCA-Temporal models the temporal covariances using the time dimension ($T$) as features and the image dimensions ($X \times Y$) as samples. 
For IFS data at low spectral resolution ($R < \num{1000}$), the wavelength dimension ($\Lambda$) can be added to the samples to leverage angular and spectral differential imaging simultaneously \citep[ASDI/CODI;][]{Kiefer2021SpectralSPHERE/IFS}. Spectral differential imaging exploits the outward motion of speckles with increasing wavelength due to chromatic PSF broadening to differentiate them from the fixed signal of the planet. 

At higher spectral resolution, where the wavelength coverage might not suffice for SDI, the signal from the planet can be disentangled from that of its host star by leveraging their significantly different spectral features.
This simple idea is implemented in different ways in the literature; however, most methods follow a similar scheme: (i) first create a stellar PSF model using expert knowledge about HCI and telluric absorption lines, and then (ii) correct any remaining systematics using PCA.
\citet{Hoeijmakers2018Medium-resolutionImaging} implemented this scheme on $K$-band SINFONI data of $\beta$~Pic~b at $R \sim \num{4000}$ and first coined the term molecular maps. Their method first built a PSF model by dividing the 3D IFS data by a master stellar spectrum extracted from the brightest pixels. The result was then low-pass filtered and multiplied by the master stellar spectrum. The role of the low-pass filter is to create a model of the wavelength-dependent stellar PSF; however, it also includes the continuum of the planet, which therefore disappears when the PSF is subtracted. Since telluric absorption lines of CH$_{4}$ remained in the residuals, they applied the SYSREM algorithm \citep{Tamuz2005CorrectingCurves,Birkby2013Detection3.2m}, which reduces to PCA if the data have constant uncertainties, in order to remove these systematics. In this framework, the features are defined along the spectral dimension ($\Lambda$) whilst the samples consist of all spaxels of a frame along the two spatial dimensions ($X \times Y$).
The same method, where SYSREM was replaced with PCA, was later applied to VLT/MUSE and SINFONI data of the PDS~70 system \citep{Haffert2019Two70,Cugno2021MolecularSystem}, where it was coined high-resolution spectral differential imaging (HRSDI).
\citet{Ruffio2019RadialSpectroscopy} introduced a forward-model framework to simultaneously fit the planet spectrum together with the stellar PSF, which is  constructed  as in \citet{Hoeijmakers2018Medium-resolutionImaging}. This approach was later improved \citep{Hoch2020Moderate-resolutionB,Ruffio2021DeepSpectroscopy} to include PCA to model the remaining systematics after an initial stellar PSF fit and subtraction. 
In this framework, the features are defined as $5\times5$ spaxel patches ($\mathcal{P} \times \Lambda$, with $\mathcal{P} \subset X \times Y$), whilst the samples are defined from the rest of the frame.
These two frameworks explicitly enforce the behaviour of the stellar PSF as a function of wavelength: the low-frequency variations come from the chromatic broadening of the stellar PSF and the continuum of the stellar spectrum, whilst the high-frequency variations are due to the telluric absorption lines imprinted on the stellar spectrum. Any remaining systematics due to wavelength calibration errors or variable telluric line strengths are subsequently identified and removed with PCA. 

For this work we  opted to leave the full  modelling of the stellar PSF to PCA (steps (i) and (ii) combined).
We defined the features along the spectral dimension ($\Lambda$) and the samples consist of all spaxels in the frame ($X \times Y$).
We computed a new PCA basis separately for each cube, meaning that the time dimension ($T$) is not used.
Afterwards, each spaxel of each cube was projected onto the first \num{250} principal components (PCs), thereby creating a model of the stellar PSF for each cube. 
The number of principal components used to build the PCA basis was determined by a blind search.
The planet can be detected with 50 PCs, but the highest S/N was obtained with 250 PCs. 
Since our approach uses PCA for the full PSF model (i and ii) and not only for the subtraction of the remaining systematics (ii) as in \citep[e.g. ][]{Hoeijmakers2018Medium-resolutionImaging,Hoch2020Moderate-resolutionB,Ruffio2021DeepSpectroscopy}, a significantly larger number of components is needed.
The PCA-based noise model was subtracted from the data resulting in a sequence of residual cubes.
These residuals were then aligned with the previously derived position of the star in each cube, and median-combined along the time dimension to produce a final residual cube.  
We refer to this technique as spectral PCA.

Although PCA has been used with features along the spectral axis ($\Lambda$) and samples along the two spatial axes ($X \times Y$) before \citep{Hoeijmakers2018Medium-resolutionImaging}, the novelty of this technique consists in using it to fully model the 3D stellar PSF. To underline the similarities and differences between some of the existing PSF subtraction techniques and Spectral PCA, we report their defining concepts in Table~\ref{tab:pca_methods}. 

\section{Results}
\label{section:results}

\subsection{Molecular maps}
\label{section:results:molmaps}

\begin{figure*}[t]
    \centering
    \includegraphics[width=\hsize]{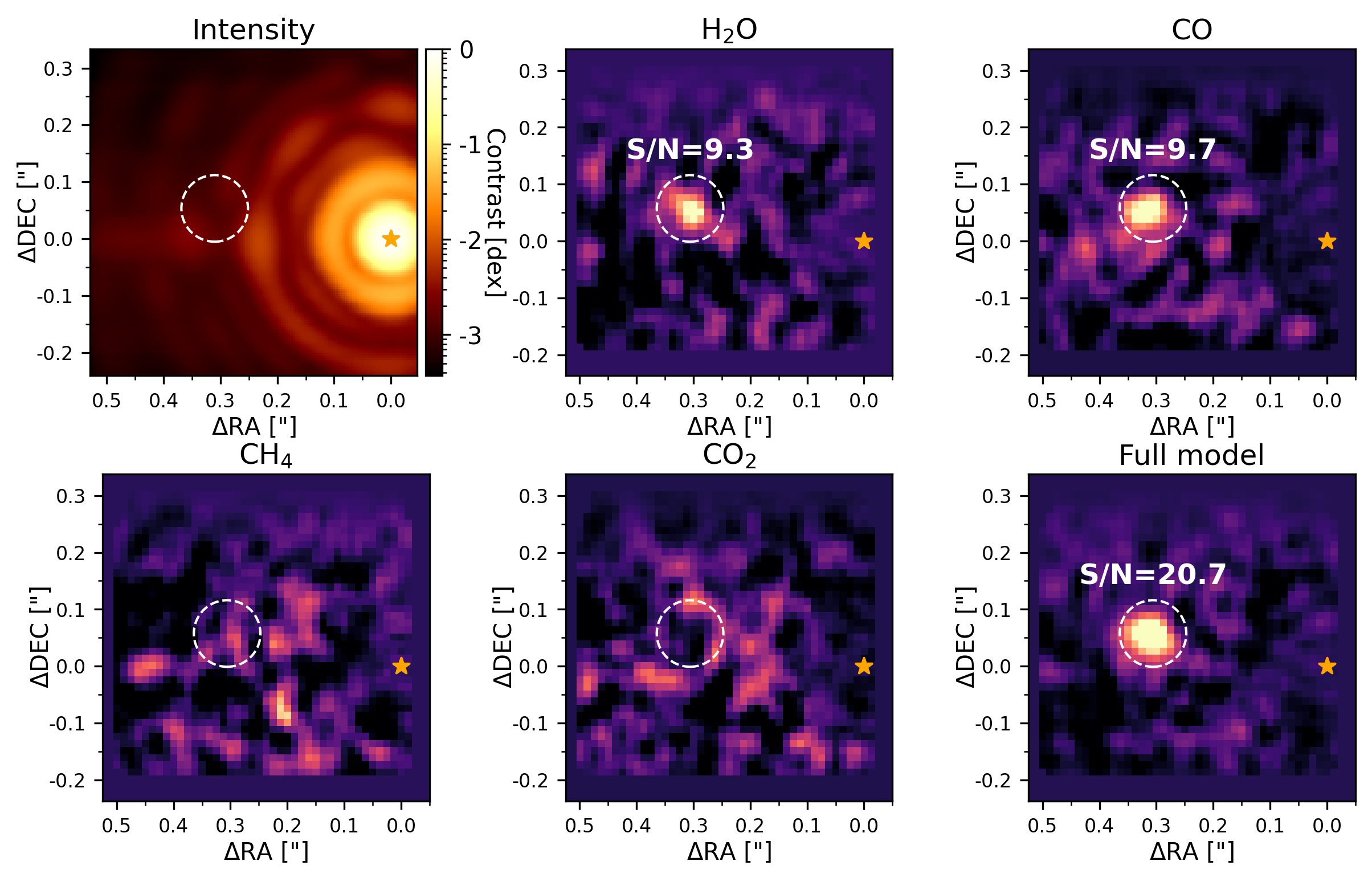}
    \caption{PSF intensity (\textit{top left}) and molecular maps of the AF~Lep system. The planet AF~Lep~b is revealed in the H$_{2}$O and CO maps, and with the full model from \citet{Balmer2025VLTI/GRAVITYMetallicity}, but not in the CH$_{4}$ and CO$_{2}$ maps. The orange star indicates the position of the star, while the white dashed circle, whose radius is equal to 1\,$\lambda/D$, indicates the position of the planet.}
    \label{fig:molmaps}
\end{figure*}

Molecular maps are obtained by cross-correlating each spaxel in the residual cube with molecular spectral templates, thereby optimally matching the molecular signature from the exoplanet atmosphere while dampening any remaining systematics. Our spectral templates were calculated using \texttt{petitRADTRANS} \citep{Molliere2019PetitRADTRANS:Retrieval} with only one molecule as a source of opacity at a time, setting its mass fraction to \qty{-2}{\dex}, and adopting the maximum a posteriori pressure--temperature profile constrained by \citet{Balmer2025VLTI/GRAVITYMetallicity}. We calculated spectral templates for H$_{2}$O, CO, CH$_{4}$, and CO$_{2}$ as they produce significant spectral features in the $K$ band. Additionally, we prepared a full-model template corresponding to the maximum a posteriori parameters derived by \citet{Balmer2025VLTI/GRAVITYMetallicity} for AF~Lep~b (i.e. $\log g = \qty{4.17}{\dex}$, $\text{Fe/H} = \num{0.74}$, $\text{C/O} = 0.50$) containing H$_{2}$O, CO, CH$_{4}$, CO$_{2}$, H$_{2}$S, HCN, NH$_{3}$, FeH, TiO, PH$_{3}$, K, VO, and Na, but ignoring the cloud treatment. We then removed the continuum of the spectral templates using a Gaussian filter with a standard deviation of \qty{2.6}{\nano\meter}. We note that the precise size of the filter does not significantly influence the molecular maps as long as the continuum is effectively removed; however, we did not systematically search for the optimal filter size to create the molecular maps. We cross-correlated the final residual cube with the spectral templates by Doppler-shifting them over a range of radial velocities from \num{-1000} to \qty{1000}{\kilo\meter\per\second} in steps of \qty{1}{\kilo\meter\per\second} using the function \texttt{crosscorrRV} from the Python package \texttt{PyAstronomy} \citep{pya}. Following \citet{Petrus2021Medium-resolutionB}, molecular maps were created by evaluating the cross-correlation function (CCF) at the RV of the planet. As an additional step, we standardised the map by subtracting the mean and dividing by the standard deviation over the whole image. 

We show the resulting molecular maps in Fig.~\ref{fig:molmaps}. The planet is clearly revealed in the H$_{2}$O, CO, and full-model maps at S/N values of \num{9.3}, \num{9.7}, and \num{20.7}, respectively. The calculations made to quantify the S/N follow \citet{Petrus2021Medium-resolutionB} with the modification to only consider values of the CCF at the same angular separation as the planet as the speckle noise is expected to depend on the angular separation \citep[see e.g.][]{Marois2008ConfidenceImaging}. More details on the computation of the S/N metric are given in Appendix~\ref{app:detection_S/N}. The signal is located at a position of $\Delta\alpha \approx +0\farcs30$ and $\Delta\delta \approx +0\farcs06$ (i.e. $\rho \approx 0\farcs3$, $\theta \approx 79^{\circ}$), as expected from previous direct imaging of the planet \citep{DeRosa2023DirectLeporis,Mesa2023AFData,Franson2023AstrometricLep,Bonse2025Use2011,Balmer2025VLTI/GRAVITYMetallicity}, and shows a (non-calibrated) RV of approximately \qty{20}{\kilo\meter\per\second}. 

\begin{figure*}[t]
    \centering
    \includegraphics[width=\hsize]{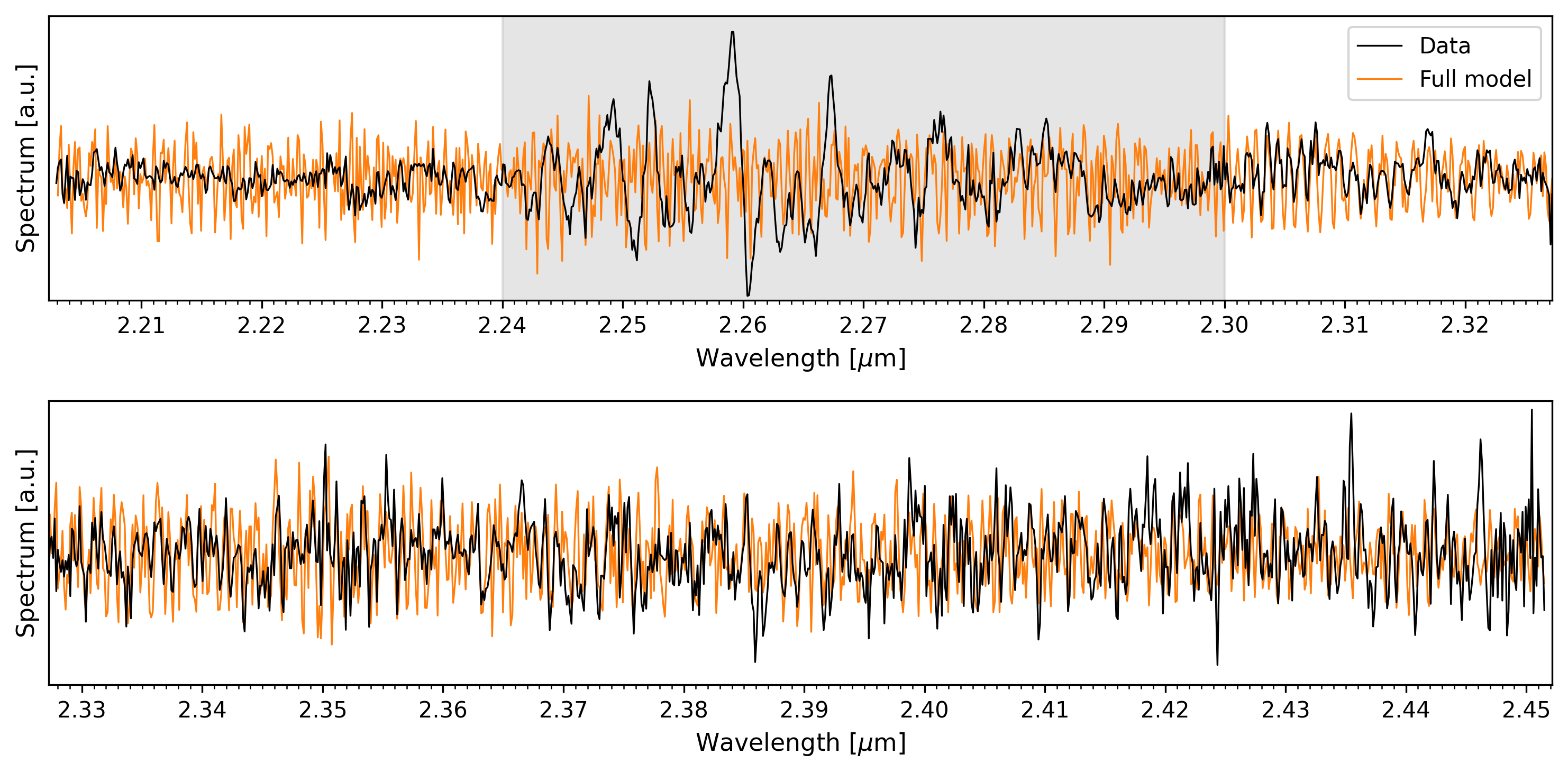}
    \caption{Spectrum of AF~Lep~b after PSF subtraction (\textit{black}) compared to our full model (\textit{blue}). The CO-dominated region between \qtyrange[range-phrase=~and~]{2.30}{2.32}{\micro\meter} matches the model particularly well despite the low S/N of the data. The shaded area represents the mask applied to the data to remove the instrument stray light described in Sect.~\ref{section:methods:data_reduction}.}
    \label{fig:final_spectrum_chem_equ}
\end{figure*}

\begin{figure}[t]
    \centering
    \includegraphics[width=\hsize]{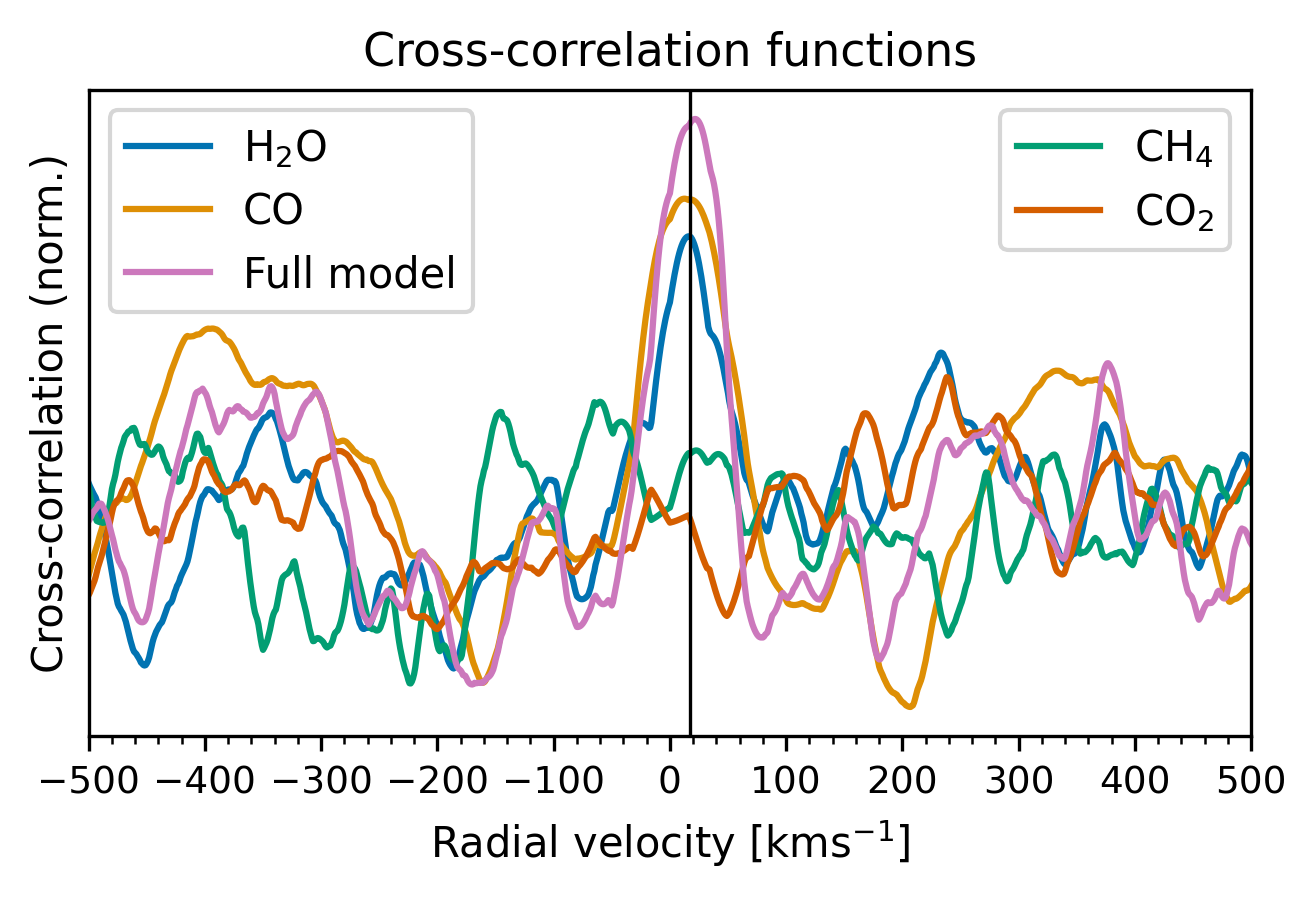}
    \caption{Cross-correlation functions with molecular templates at the position of the planet in Fig.~\ref{fig:molmaps}. The peaks indicate the presence of H$_{2}$O and CO in the atmosphere of AF~Lep~b at an (uncalibrated) RV of approximately \qty{20}{\kilo\meter\per\second} (\textit{vertical line}). The cross-correlation functions were standardised over the spatial axes in the same way as our molecular maps:  the cross-correlation map at each RV was mean-subtracted and normalised by its standard deviation over the whole image.\looseness=-1}
    \label{fig:ccf}
\end{figure}

The extracted spectrum of the planet is shown in Fig.~\ref{fig:final_spectrum_chem_equ} together with our full model template for reference. Although the planet spectrum has a low S/N and might still contain residual stellar light, absorption lines are nevertheless well resolved, for example in the region from \qtyrange{2.30}{2.32}{\micro\meter} where CO dominates the spectrum. The stray light mentioned in Sect.~\ref{section:methods:data_reduction} can be seen between \qtyrange[range-phrase=~and~]{2.24}{2.30}{\micro\meter}. The spectral templates used for the cross-correlation are shown in Fig.~\ref{fig:spectrum} together with the spectra at the position of the planet before and after PSF subtraction and aperture photometry. Finally, we show the cross-correlation functions with the different molecules at the position of the planet in Fig.~\ref{fig:ccf}. 
\looseness=-1

\subsection{Radial velocity}
\label{section:results:RV}

To measure the RV from the cross-correlation function, we would like to interpret it as the position of the peak in the cross-correlation function. However, each frame might have slightly different systematics. Errors in the wavelength solution, imperfect PSF subtraction, and  stray light, for example, might affect the shape of the cross-correlation function as well as the position of its peak. Since the planet is not detected in single frames, we adopted a bootstrapping approach \citep{BeranAnBootstrap} to replicate the observation many times by randomly sampling the frames and combining them, thereby quantifying the random effect of the instrument and processing the systematics. Furthermore, we corrected for the relative motion of the Paranal observatory with respect to the AF~Lep system.
Since we detected the planet with three different spectral templates, we can measure its RV with each model. 
The exact details of the measurement and calibration of the RV of AF~Lep~b are given in Appendix~\ref{app:RV}.
We report the following results for each model: for H$_{2}$O, we obtain $\Delta v_{\mathrm{R,P}\star}^{(\mathrm{H_{2}O})} =\qty{8.9+-2.5}{\kilo\meter\per\second}$; for CO, we obtain $\Delta v_{\mathrm{R,P}\star}^{(\mathrm{CO})} =\qty{6.1+-2.8}{\kilo\meter\per\second}$; and for the full model, we obtain  $\Delta v_{\mathrm{R,P}\star}^{(\mathrm{Full})} =\qty{7.8+-1.7}{\kilo\meter\per\second}$. These three values are consistent with each other and allow the orbital solution to be constrained. We note that since the full model contains H$_{2}$O and CO, the three measurements are not independent of each other, which might explain why the RV value obtained with the full model lies between the other two values. This also makes the combination of the three values non-trivial. Instead, we chose the measurement resulting in the highest S/N detection of AF~Lep~b (i.e. the full model), and take it as our final RV measurement $\Delta v_{\mathrm{R,P}\star} = \Delta v_{\mathrm{R,P}\star}^{(\mathrm{Full})} =\qty{7.8+-1.7}{\kilo\meter\per\second}$.

For a planet that has been discovered simultaneously by three independent teams, it is not surprising that its RV has also been measured by another instrument at a time similar to that of our ERIS observations. \citet{Denis2025CharacterisationVLT/HiRISE} have also observed AF~Lep~b with the new VLT/HiRISE \citep{Vigan2024FirstExoplanets} instrument at high spectral resolution ($R \sim \num{100000}$). Their analysis delivers the value $\qty{10.51+-1.03}{\kilo\meter\per\second}$ \citep{Denis2025CharacterisationVLT/HiRISE}. Our final RV measurement lies within $1.6\,\sigma$ of their result.

To validate the uncertainty delivered by our bootstrapping approach, we ran a simple simulation using the exposure time calculator of ERIS/SPIFFIER.\footnote{The ESO exposure time calculator for ERIS is available at the following link: \url{https://etc.eso.org/eris}.} For a field object with an effective temperature of \qty{800}{\kelvin} and an apparent magnitude of \qty{16.77}{\magnitude} such as AF~Lep~b,  the measured spectrum is estimated to have a $\text{S/N} \approx 10$ for an identical observing set-up,  as described in Sect.~\ref{section:methods:observations}. We simulate \num{1e4} noisy spectra at this S/N using our full model of the planet and subsequently measure the RV using the same cross-correlation technique as described in Appendix~\ref{app:RV} (i.e. by measuring the position of the peak of the cross-correlation function evaluated with our full model spectral template). The resulting RV measurements have a standard deviation of \qty{0.1}{\kilo\meter\per\second}. This reflects the uncertainty with which the RV can be measured for this experiment. Since the S/N calculated with the ETC ignores residual systematics from the subtraction of the stellar PSF as well as photon noise from the stellar PSF overlapping with AF~Lep~b, the S/N is expected to be much smaller. By repeating the experiment with $\text{S/N} = 1$, the resulting uncertainty is \qty{1.1}{\kilo\meter\per\second}, which is already much closer to our uncertainty of \qty{1.7}{\kilo\meter\per\second}. The actual S/N of our extracted spectrum is difficult to estimate, especially considering that the continuum is lost during PSF subtraction. However, since CO absorption lines are visible by eye, the S/N cannot be far from \num{1}.
Additionally, as we describe in Sect.~\ref{section:methods:wavelength}, measuring the wavelength error after applying our pipeline showed a residual error of \qty{0.8+-0.3}{\kilo\meter\per\second} over the whole observation (i.e. a bias of \qty{0.8}{\kilo\meter\per\second} and a random scatter of \qty{0.3}{\kilo\meter\per\second}). Assuming that the bias can be corrected perfectly (see Appendix~\ref{app:RV} for details on the correction of this bias), the random errors arising from the wavelength calibration add another uncertainty of the order of \qty{0.3}{\kilo\meter\per\second}. Our actual uncertainty of \qty{1.7}{\kilo\meter\per\second} can therefore be quite well explained as a combination of residual wavelength errors and low S/N spectrum.
\looseness=-1

\section{Discussion}
\label{section:discussion}

\subsection{Orbital constraint}
\label{section:discussion:orbital_constraint}

Our RV measurement can be used to remove the ambiguity in the previously constrained orbital solution of AF~Lep~b \citep{DeRosa2023DirectLeporis,Franson2023AstrometricLep,Mesa2023AFData,Bonse2025Use2011,Zhang2023ELementalAtmospheres,Balmer2025VLTI/GRAVITYMetallicity}.
The orbital fit obtained by \citet{Balmer2025VLTI/GRAVITYMetallicity} on all previous imaging, astrometry, stellar RV, and GRAVITY data can be evaluated to predict the RV of the companion relative to its host star on the night of our ERIS observations. Due to the bimodality of the orbital solution, there are two equiprobable radial velocities: (a) $\Delta v_{\mathrm{R,P\star}}=\qty{10+-2}{\kilo\meter\per\second}$ associated with the argument of periapsis $\omega=109^{+13}_{-21}$~deg and longitude of the ascending node $\Omega=248.8^{+0.4}_{-0.7}$~deg, and (b) $\Delta v_{\mathrm{R,P\star}}=\qty{-10+-2}{\kilo\meter\per\second}$ associated with the other solution $\omega=289^{+13}_{-21}$~deg, $\Omega=68.8^{+0.4}_{-0.7}$~deg (the other orbital parameters are all equal in both cases). Here, the ascending node $\Omega$ is defined in the same way as in \texttt{orvara} \citep{Brandt2021TheEdition}; in other words,  it denotes the position angle at which the planet crosses the sky plane moving towards the observer \citep[we note that \texttt{orbitize!} uses the opposite definition;][]{Blunt2024OrbitizeCommunity}. 
\looseness=-1

Our RV measurement using the full model template:  $\Delta v_{\mathrm{R,P}\star} = \qty{7.8+-1.7}{\kilo\meter\per\second}$. It is consistent with the first orbital solution within $1.3\,\sigma$, while excluding the other at $11\,\sigma$. 
Our measurement therefore shows that the orbital solution $\omega=109^{+13}_{-21}$~deg, $\Omega=248.8^{+0.4}_{-0.7}$~deg is the correct one. This means that (i)~the relative RV of AF~Lep~b was close to its maximum value at the epoch of our observations, (ii)~the planet recently crossed the sky plane moving behind the star (i.e. away from the Earth), and (iii)~more generally the planet is located behind the star (from the point of view of the Earth) between the position angles $68.8^{+0.4}_{-0.7}$~deg and $248.8^{+0.4}_{-0.7}$~deg.
In particular, it is now possible to compute a phase curve of AF~Lep~b, which is crucial for the planning of follow-up observations in visible light. Moreover, this delivers one of the two puzzle pieces necessary to determine the true obliquity of AF~Lep~b (i.e. the angle between its orbital angular momentum and the spin axis of the star). The other piece, the orientation of the spin axis of the star, can be determined by measuring the displacement of the photocentre  due to stellar rotation \citep{LeBouquin2009TheInterferometry,Kraus2020SpinOrbitSystem}.

\subsection{Implications for the atmospheric chemistry of AF~Lep~b}
\label{section:discussion:chemistry}

In itself, our detection of H$_{2}$O and CO in the atmosphere of AF~Lep~b is not surprising as these molecules have already been found in similar objects. Several atmospheric studies have started to directly constrain the mass fractions of H$_{2}$O, CO, and CH$_{4}$ in such objects \citep{Zhang2023ELementalAtmospheres,Palma-Bifani2024AtmosphericModeling,Balmer2025VLTI/GRAVITYMetallicity}. Chemical disequilibrium, which is driven by vertical mixing, has been advanced as a potential explanation for the high abundance of CO found by these studies. At an effective temperature of \qty{800}{\kelvin}, CH$_{4}$ is expected to be chemically favoured over CO \citep{Molliere2022InterpretingAssumptions}, hence the need for an  upward stream of CO from hotter deeper layers. Therefore, our detection of H$_{2}$O and CO together with our non-detection of CH$_{4}$ are compatible with the chemical disequilibrium hypothesis. To investigate whether our new ERIS observations also favour this hypothesis over chemical equilibrium, one would need to constrain the molecular abundances of H$_{2}$O and CO. This requires the use of a dedicated atmospheric retrieval framework such as CROCODILE \citep{Hayoz2023CROCODILE} and will be investigated in a future article.

\subsection{Stellar and telluric contamination}
\label{section:discussion:stellar_contamination}

The stellar PSF is still very strong at the position of the planet: around $10^{-3.25}$ times its maximum value. This means that the planet with its contrast of $\Delta K = \qty{11.84}{\magnitude}$ \citep{DeRosa2023DirectLeporis} is around 30 times fainter than the PSF at its position. There is therefore a risk of contamination of the planet signal by the stellar spectrum. More precisely, since the spectrum of AF~Lep only contains atomic lines and at most very weak molecular lines, the risk instead comes from the strong telluric absorption lines imprinted in the stellar spectrum by the atmosphere of the Earth.
However, molecular maps remedy this problem by measuring the cross-correlation at the position of the planet and everywhere else in the field of view. Therefore, if the spectral templates correlate with the stellar spectrum, then this contamination can be measured as a cross-correlation signal in a different place than the planet. This is the reason for our chosen S/N metric, which ensures that any contamination by the tellurics-imprinted stellar spectrum is taken into account.
\looseness=-1

\subsection{High-contrast capabilities of ERIS}
\label{section:discussion:contrast_performance}

\begin{figure*}[ht]
\centering
    \includegraphics[width=0.33\linewidth]{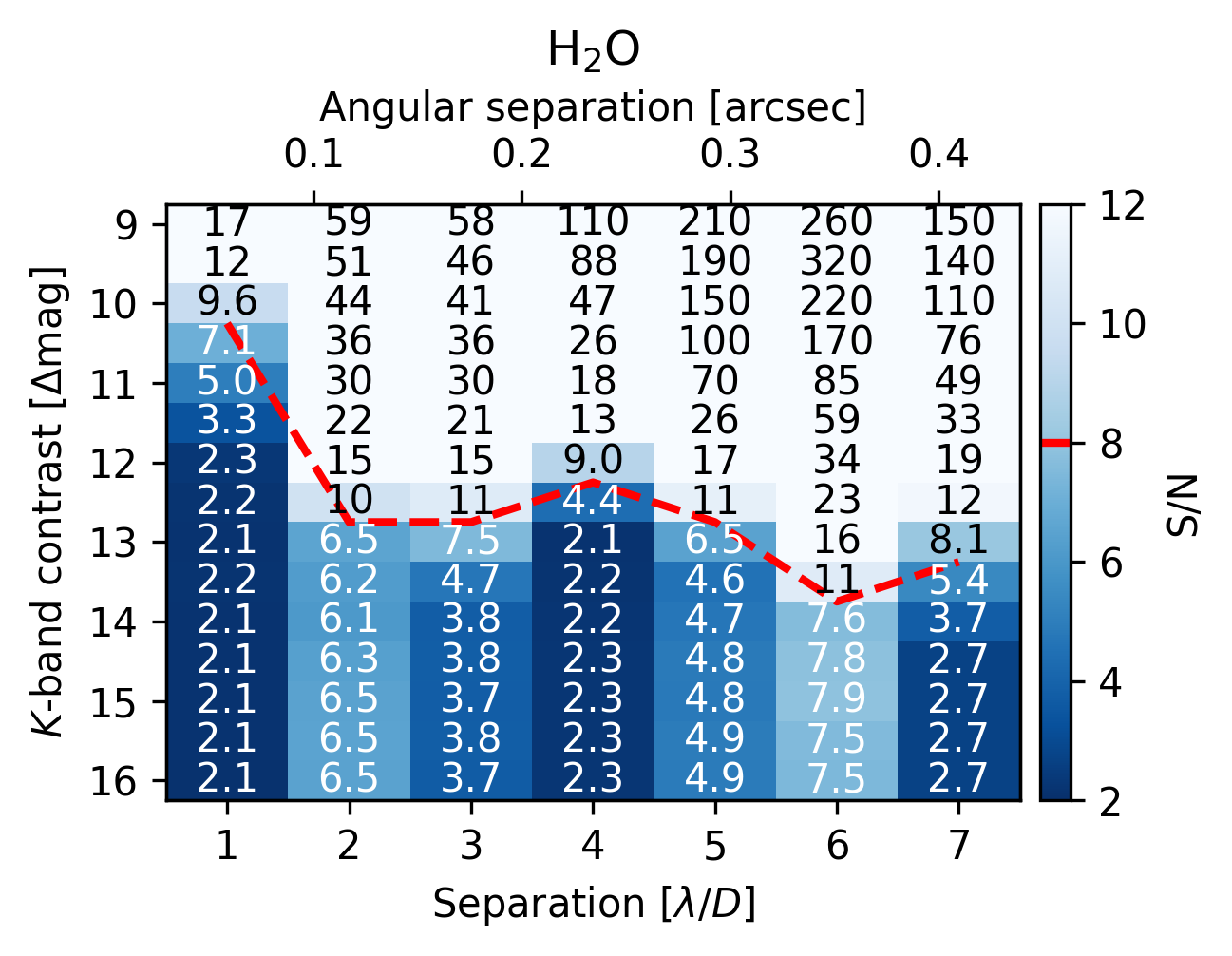}
    \includegraphics[width=0.33\linewidth]{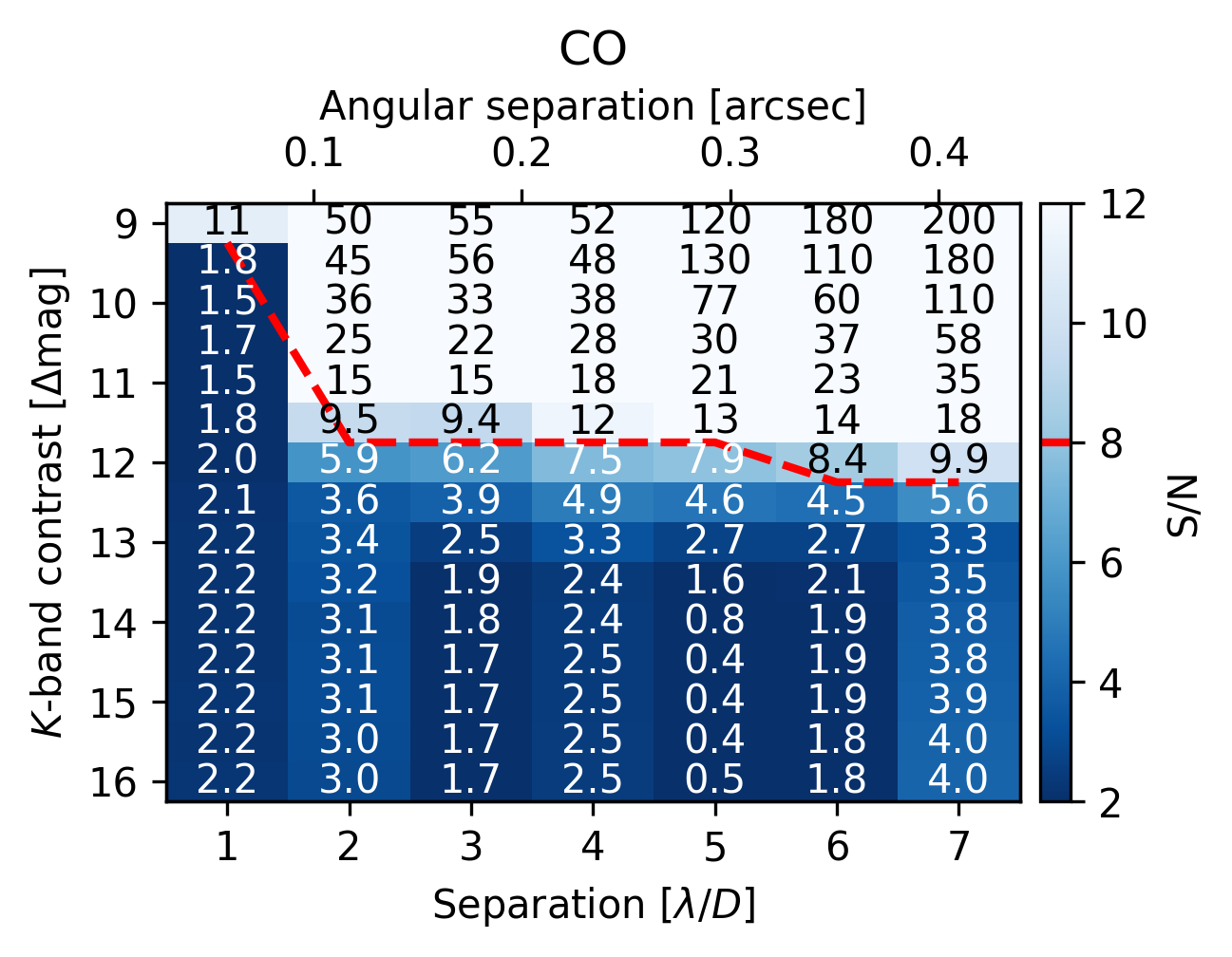}
    \includegraphics[width=0.33\linewidth]{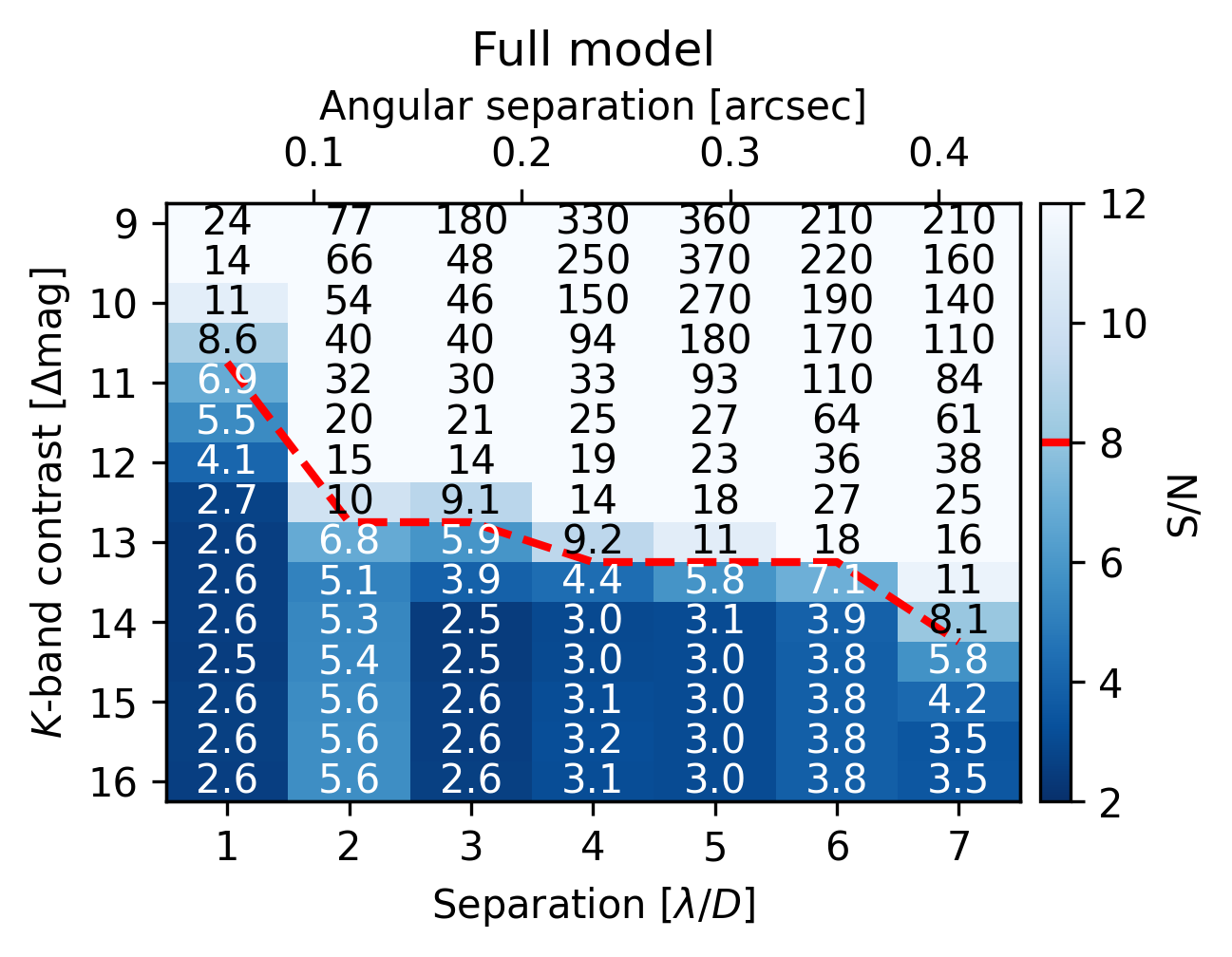}
\caption{Contrast grids computed by injecting fake planets with our full model spectrum from Sect.~\ref{section:results:molmaps} at different contrast and angular separation from the host star and retrieving them in the molecular maps of H$_{2}$O (\textit{left}), CO (\textit{centre}), and with the full model (\textit{right}). The dashed red line corresponds to a S/N of 8, which we use as the detection threshold to compute our detection limits.}
\label{fig:contrast_grid}
\end{figure*}

The high significance of our detection of AF~Lep~b hints at deeper detection limits. 
Establishing a robust detection limit for molecular mapping is more complicated than for imaging, due to the high dependence on the atmospheric properties of the companions and the spectral type of the host star. The pressure--temperature profile, the chemical composition of the atmosphere, and the presence and composition of clouds strongly affect the shape and strength of the absorption lines, which in turn will affect the signal picked up by cross-correlation with spectral templates \citep{Bidot2024ExoplanetSpectro-imaging}. At the same time, the photosphere of the host star can (depending on its spectral type) contain the same molecules of interest as the companion, which might affect the efficacy of the PSF subtraction and thereby the detection limits, hindering the generalisation to another stellar system \citep{Bidot2024ExoplanetSpectro-imaging}. Accounting for all these factors is beyond the scope of this paper.
Nevertheless, we wanted to provide indicative detection limits that can be used as a reference point for the planning of future observations with ERIS/SPIFFIER. To do so, we set up a fake planet injection and retrieval test as is customary in HCI \citep[see e.g.][]{Bonse2023ComparingNoise}, but we constrained the analysis to the detection of companions spectrally identical to AF~Lep~b.
Concretely, we used our full model from Sect.~\ref{section:results:molmaps} as planet spectrum.

We created a PSF reference by considering the six frames where the star is closest to the centre of the field of view, cropping them to a size of $0\farcs3 \times 0\farcs3$, aligning them, and finally median-combining them. To inject a fake planet at a given contrast, we divided the datacube of the PSF reference by the stellar spectrum and multiplied it by the planet spectrum and the contrast. The stellar and planet spectra were normalised using synthetic photometry with the filter SPHERE/IRDIS K12\_2, as it was used to report the contrast of AF~Lep~b in the $K$ band \citep{DeRosa2023DirectLeporis}. For the stellar spectrum, we used a BT-NextGen model \citep{Allard2012ModelsExoplanets} evaluated at the grid point closest to the stellar parameters reported by the \textit{Gaia} DR3 (i.e. $T_{\mathrm{eff}} = \qty{5900}{\kelvin}$, $\log g = \qty{4}{\dex}$, $\text{Fe/H} = \qty{-0.5}{\dex}$).
Fake planets were injected into each datacube of the reduced data at a fixed relative position with respect to the host star and with no Doppler shift. We injected the fake planets right before the PSF subtraction and subsequently applied the rest of the pipeline to the fiducial datasets.

Figure~\ref{fig:molmaps} shows that there is not a lot of room in the $0\farcs8 \times 0\farcs8$ field of view of ERIS/SPIFFIER to inject fake planets around the host star. This forced us to inject planets at only one position angle, namely at $\qty{105}{\deg}$ (as measured from north through east), and only up to 0\farcs4 angular separation. Concretely, we used a range of angular separation between 1 and 7~$\lambda/D$ in steps of 1~$\lambda/D$ 9 (i.e. between 0\farcs06 and 0\farcs41).
For the contrast, we selected a range between 9 and 16~$\Delta$mag in steps of 0.5~$\Delta$mag. This allowed us to create a contrast grid similar to the ones computed by \texttt{applefy} \citep{Bonse2023ComparingNoise}.
\looseness=-1

\begin{figure}[t]
    \centering
    \includegraphics[width=\hsize]{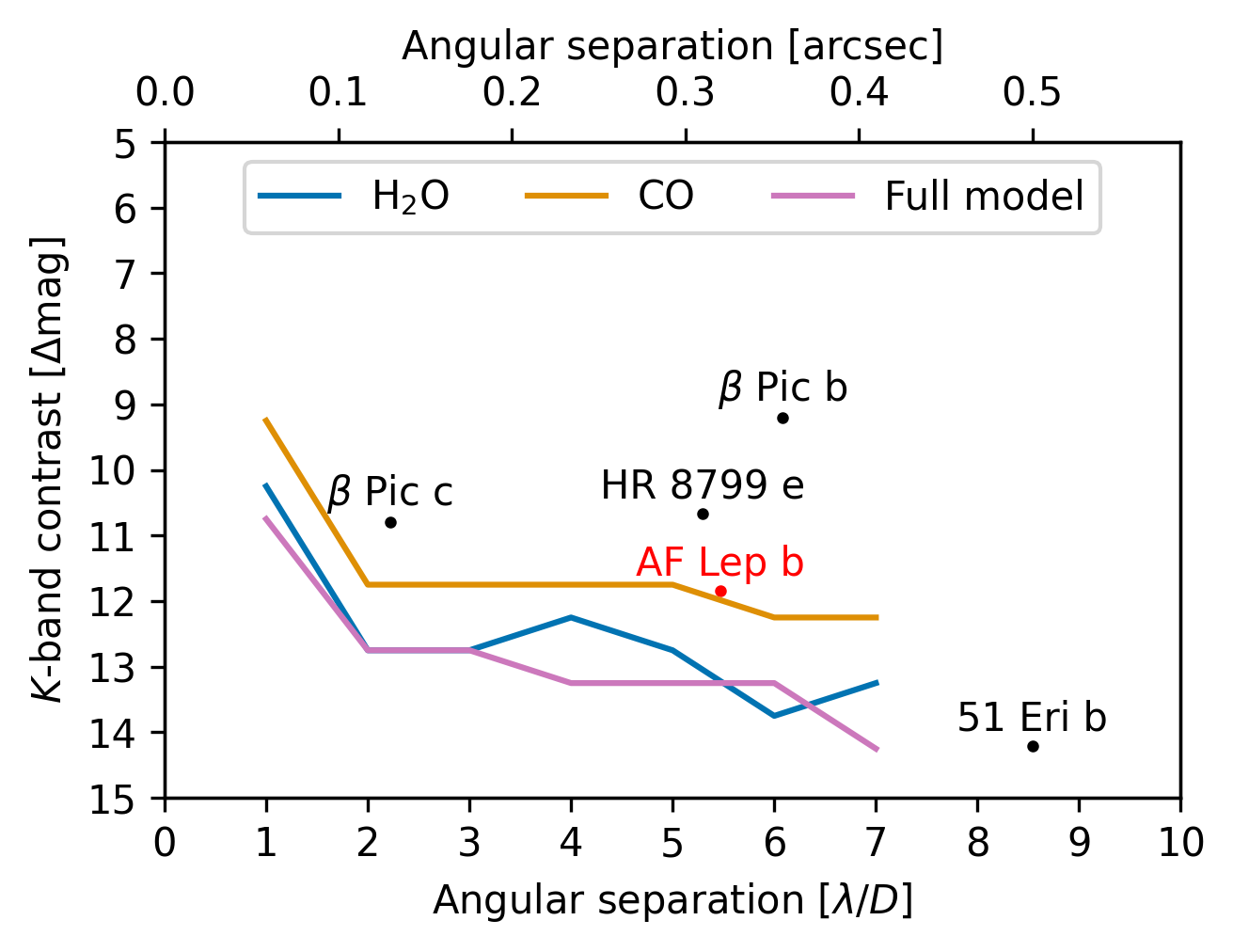}
    \caption{Detection limits at $\text{S/N} = 8$ for H$_{2}$O, CO, and the full model for our ERIS/SPIFFIER observations of the AF~Lep system. The detection limits were calculated for the full \qty{178}{\minute} total integration time, which required just over \qty{4}{\hour} of telescope time, and did not use ADI as it was not yet available at the time of our observations. The contrast and angular separations of the other directly imaged exoplanets indicated in this figure are taken from the following studies: $\beta$~Pic~c from \citet{Nowak2020DirectC}, $\beta$~Pic~b from \citet{Bonnefoy2011HighM}, HR~8799~e from \citet{Marois2010Images8799}, AF~Lep~b from \citet{DeRosa2023DirectLeporis}, and 51~Eri~b from \citet{Samland2017SpectralVLT/SPHERE}.
    \looseness=-1}
    \label{fig:contrast_curve}
\end{figure}

The resulting contrast grids for the molecular maps of H$_{2}$O, CO, and for the full model are shown in Fig.~\ref{fig:contrast_grid}. From these grids, we extracted contrast curves by setting the detection threshold to $\text{S/N} = 8$ to enforce a clear detection.
We show the three extracted $\text{S/N} = 8$ detection limits in Fig.~\ref{fig:contrast_curve} together with the $K$-band contrast and angular separation of other directly imaged exoplanets as reference. To illustrate these contrast grids and our detection limits, we show a few resulting molecular maps for a sample of contrast and angular separation close to our detection limits in Fig.~\ref{fig:fake_planets}. The injected fake planets are indeed clearly detected at $\text{S/N} \geq 8$, but not necessarily at lower values, which validates our selected detection threshold.

These detection limits imply that we could have detected AF~Lep~b even if it had been at an angular separation of 2\,$\lambda/D$ (i.e. 0\farcs12). The second row of Fig.~\ref{fig:fake_planets} shows that the fake planet is retrieved at S/N of \num{15.2} and \num{14.6} with the H$_{2}$O template and the full model. This corresponds to the contrast and angular separation close to which $\beta$~Pic~c was detected using interferometry with GRAVITY \citep{Nowak2020DirectC}. The close-in exoplanet has not been detected yet with non-interferometric observations.  $\beta$~Pic~c certainly has different atmospheric properties than AF~Lep~b, which is not accounted for in this analysis. To validate the high-contrast potential of molecular mapping with ERIS/SPIFFIER in the 1--2~$\lambda/D$ range, one would have to actually detect $\beta$~Pic~c. If possible, this would significantly extend the parameter space accessible to characterise exoplanets at moderately high spectral resolution towards smaller angular separation. Molecular mapping with ERIS/SPIFFIER might even allow us to efficiently search for other young and bright super-Jovian planets that are closer in.

We would like to emphasise that these detection limits are only indicative and should not be interpreted as a guarantee to detect a companion with a contrast and angular separation located close to the contrast curves. Many assumptions went into computing them, which is why we would like to emphasise them here. The input spectrum is a simplified model of the spectrum of AF~Lep~b that does not include any treatment of clouds. The wavelength is assumed to be perfectly calibrated. The S/N metric used here \citep[cf.][]{Petrus2021Medium-resolutionB} does not take into account effects from small sample statistics and might overestimate the confidence of the detection. For each contrast and angular separation, the fake planets were only injected at one position angle, which can create a bias depending on the position of residual speckles. Finally, the contrast curves are calculated with respect to AF~Lep specifically, which means that a brighter star might result in a worse detection limit at close separations due to the presence of brighter speckles, whereas observations of a fainter star might be limited by the sensitivity of ERIS/SPIFFIER. Taking all that into account, our detection limits indicate that a similar observing set-up is likely to result in the detection of an exoplanet with similar atmospheric properties as AF~Lep~b with a contrast and angular separation that would locate it sufficiently above the detection limits shown in Fig.~\ref{fig:contrast_curve}.
\looseness=-1

\section{Summary and conclusion}
\label{section:conclusion}

In this article we demonstrated the first high-contrast spectroscopic application of the new ERIS/SPIFFIER instrument at the VLT. We observed the AF~Lep system at moderately high spectral resolution ($R \sim \num{11000}$) in the $K$ band as part of the ERIS GTO programme  allocated to ETH Zurich. Since high-contrast spectroscopic applications are not covered by the standard SPIFFIER pipeline, we developed a robust method to calibrate the wavelength solution by cross-correlating the strong telluric absorption lines imprinted on the stellar spectrum with a SkyCalc template \citep{Noll2012AnParanal,Jones2013AnParanal} and modelling the slight spectral curvature caused by the single diffraction grating within SPIFFIER. Our custom pipeline additionally corrects high-order effects by fitting a spline to the wavelength error. Furthermore, we presented spectral PCA, an innovative application of PCA to model the 3D stellar PSF and subtract it from the data.
Our findings are summarised in the following points.

\begin{itemize}
    \item The new ERIS instrument at the Very Large Telescope can detect molecules in the atmosphere of a planet with a contrast of $\Delta K = \qty{11.84}{\magnitude}$ \citep{DeRosa2023DirectLeporis} and an apparent separation of $0\farcs32$, and can be used to measure its RV with an uncertainty of \qty{1.7}{\kilo\meter\per\second}. This demonstrates that ERIS/SPIFFIER is a competitive instrument for the spectral and orbital characterisation of directly imaged exoplanets at high contrast and small angular separation. 
    \item We directly detected water and carbon monoxide in the atmosphere of AF~Lep~b at high S/N. Together with our non-detection of methane, this is compatible with the hypothesis of chemical disequilibrium currently favoured by previous observations. We will further investigate the atmospheric properties of AF~Lep~b using atmospheric retrievals in a follow-up article together with other targets observed as part of the ERIS GTO programme.\looseness=-1
    \item We measured the radial velocity of AF~Lep~b relative to its host star: $\Delta v_{\mathrm{R,P}\star} = \qty{7.8+-1.7}{\kilo\meter\per\second}$. This value indicates that the orbital solution with argument of periapsis $\omega = 109^{+13}_{-21}$~deg and longitude of ascending node $\Omega = 248.8^{+0.4}_{-0.7}$~deg is the correct one, removing the bimodal distribution found by \citet{Balmer2025VLTI/GRAVITYMetallicity}. As explained in Sect.~\ref{section:discussion:orbital_constraint}, this now allows the computation of the phase curve of AF~Lep~b necessary for the planning of follow-up observations in visible light, and delivers one piece of the puzzle necessary for the measurement of its true obliquity.\looseness=-1
    \item We computed the detection limits for our observations by injecting fake planets at different contrast and angular separations from the host star. 
    These detection limits were derived with many assumptions listed in Sect.~\ref{section:discussion:contrast_performance}, the most critical being that we used a simple model of the spectrum of AF~Lep~b, whilst the sensitivity of molecular mapping might strongly depend on the atmospheric parameters (i.e. pressure--temperature profile, chemistry, clouds). 
    Taken at face value, our detection limits seem to indicate that ERIS/SPIFFIER might be able to reach a contrast of $\Delta K = \qty{12}{\magnitude}$ at an angular separation of \qty{2}{\lambda/D} (i.e. 0\farcs12). If confirmed, this would significantly extend the parameter space available for the search and investigation of the atmospheric and orbital properties of low-mass companions and young exoplanets at moderately-high spectral resolution, as to date the only known exoplanet in this regime is $\beta$~Pic~c, which has only been directly imaged using interferometry with the VLTI/GRAVITY instrument \citep{Nowak2020DirectC}.
    \item Finally, ADI \citep{Marois2006AngularTechnique} is not strictly necessary for the detection and characterisation of exoplanets when observed at high enough spectral resolution. Pupil-tracking had not yet been commissioned for ERIS/SPIFFIER at the time of our observations. Our results demonstrate that stellar speckles can be effectively disentangled from the signal of the companion using PSF subtraction via spectral PCA.
    Future work should investigate whether molecular mapping can be combined with ADI to reach even deeper detection limits. However, if ADI is not necessary, it might relax some functionality requirements under investigation for the design of the next generation of high-contrast facilities aiming at characterising terrestrial exoplanets, such as the Planetary Camera and Spectrograph \citep[PCS,][]{Kasper2021PCSELT} at the upcoming Extremely Large Telescope.\looseness=-1
\end{itemize}

\begin{acknowledgements}
We thank the anonymous referee for their constructive feedback which helped us to greatly improve our article. JH and SPQ gratefully acknowledge the financial support from the Swiss National Science Foundation (SNSF) under project grant number 200020\_200399. Part of this work has been carried out within the framework of the National Centre of Competence in Research PlanetS supported by the Swiss National Science Foundation under grants 51NF40$\_$182901 and 51NF40$\_$205606. GC thanks the
Swiss National Science Foundation for financial support
under grant numbers P500PT\_206785 and P5R5PT\_225479. This project received funding from the European Research Council (ERC) under the European Union’s Horizon 2020 research and innovation programme (grant agreement No 819155).\\
\textit{Contributions}: JH proposed the project, wrote the observing proposal, prepared and analysed the observations, and wrote the article. MJB provided insightful discussions on PSF subtraction techniques and helped writing the corresponding section. AAB, YC, and RD provided help with the implementation of our custom wavelength calibration. MJB, FD, and HMS provided insightful discussions about orbital fitting. The first 16 co-authors listed after JH all provided useful feedback during the project or while writing the manuscript. All other co-authors are included for their past contributions to designing, building, and commissioning the ERIS instrument, and are ordered alphabetically.
\end{acknowledgements}

{\tiny
\bibliographystyle{aa}
\bibliography{bibliography.bib}
}

\begin{appendix}

\onecolumn
\section{Wavelength calibration}
\label{app:wavelength_calibration}

In this appendix, we give a full description of our wavelength calibration procedure as summarised in Sect.~\ref{section:methods:wavelength}. We then estimate its accuracy and indicate the average wavelength correction applied as a function of time during the whole observation. As already stated in the main text, our custom wavelength calibration consists of three steps, which we extensively describe in the following.

\begin{figure*}[ht]
\centering
    \begin{subfigure}{\textwidth}
        \centering
        \includegraphics[width=0.95\linewidth]{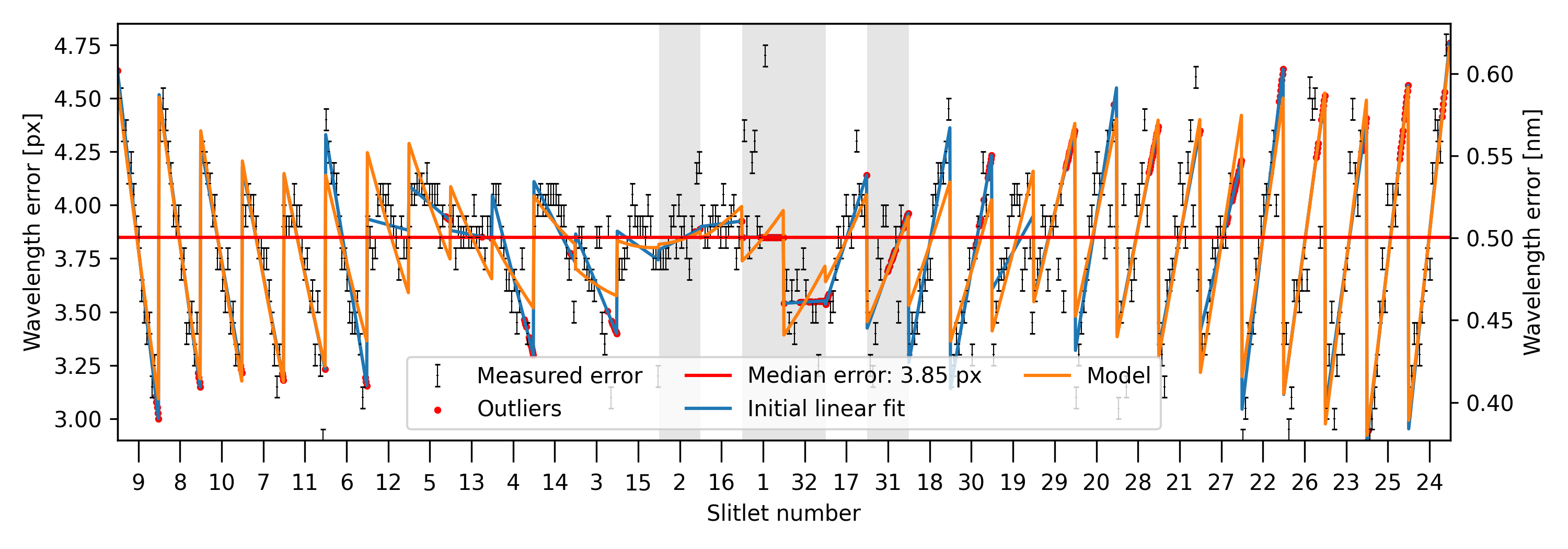}
        \caption{}
        \label{fig:wavelength_calibration:across_slit}
    \end{subfigure}

    \begin{subfigure}{0.49\textwidth}
        \centering
        \hfill\includegraphics[width=0.95\textwidth]{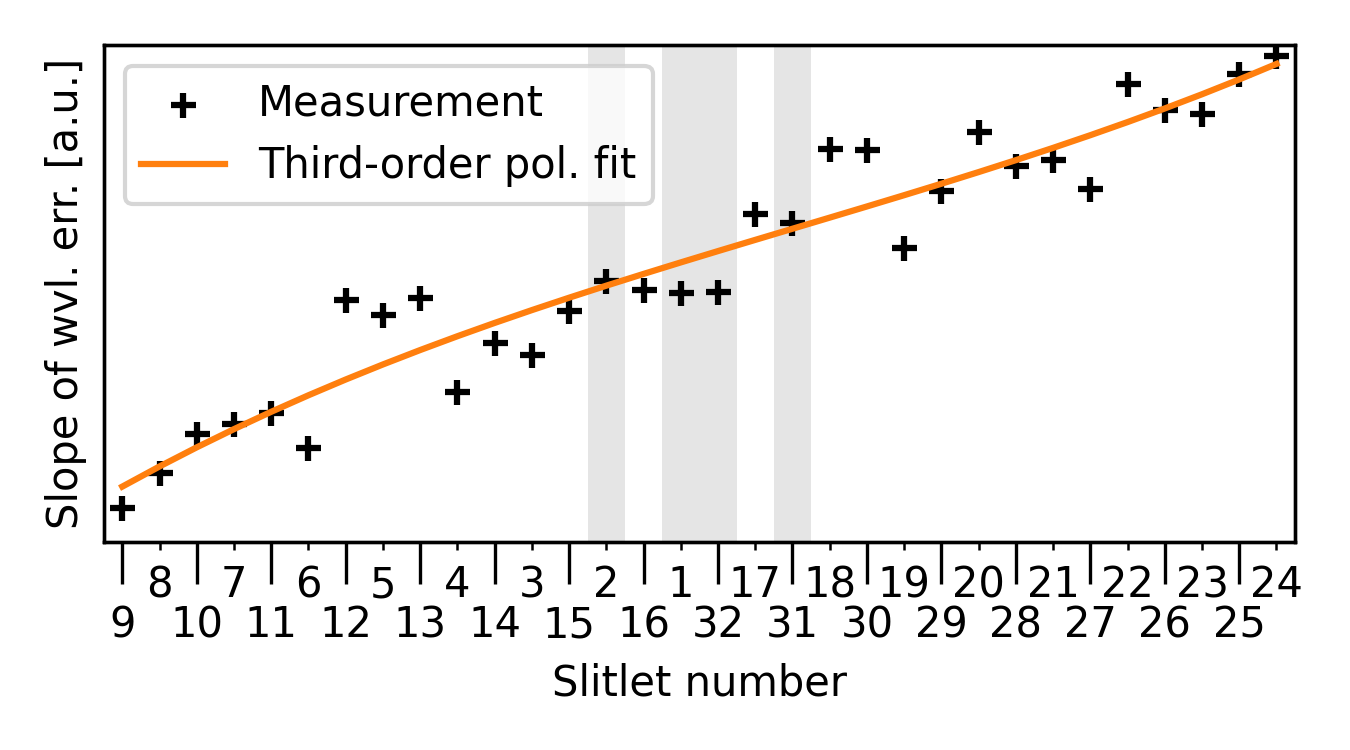}
        \caption{}
        \label{fig:wavelength_calibration:third_order}
    \end{subfigure}
    \begin{subfigure}{0.49\textwidth}
        \centering
        \includegraphics[width=0.95\textwidth]{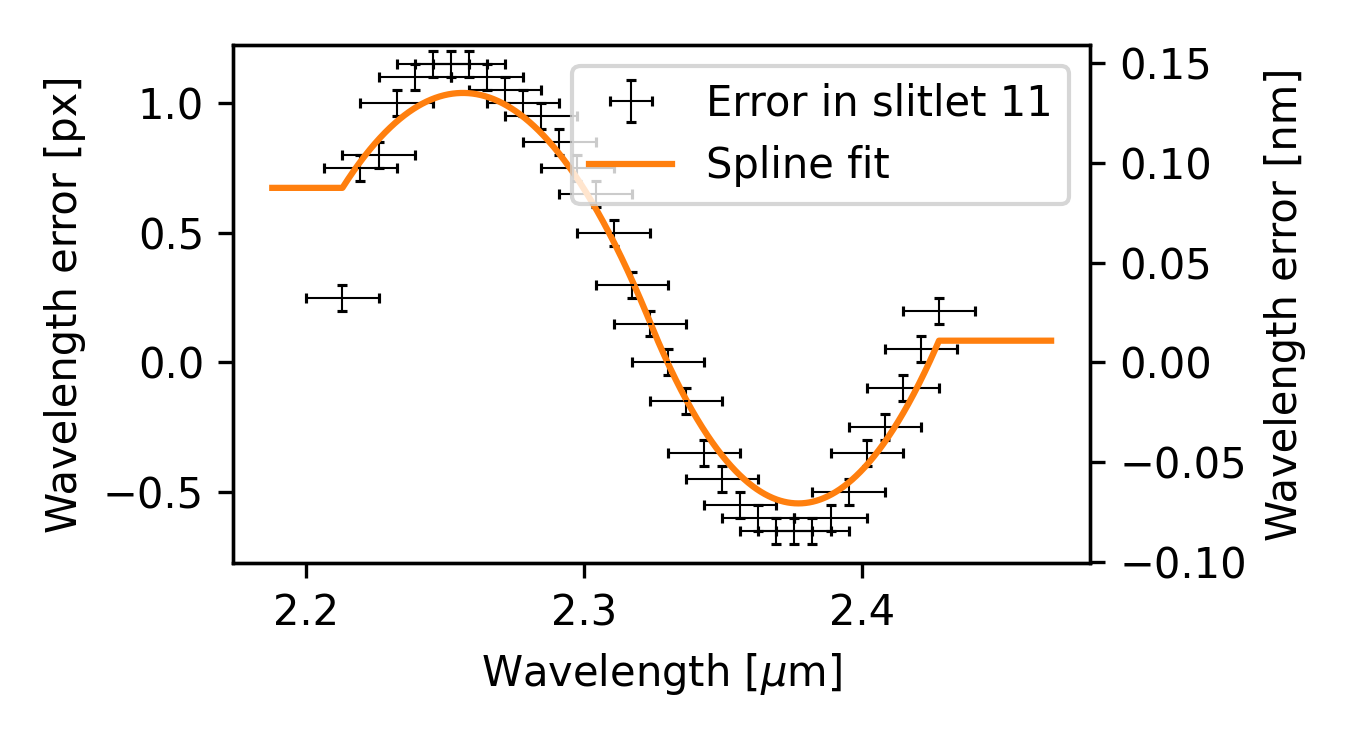}\hfill
        \caption{}
        \label{fig:wavelength_calibration:spline_fit}
    \end{subfigure}
\caption{Illustration of the steps involved in our custom wavelength calibration applied to an arbitrary frame. Panel (a): Measurement of the wavelength error across all slitlets by cross-correlating each spaxel with a telluric transmission template (black). We only show every fourth measurement for readability. Bad measurements are indicated by red dots, while the red line indicates the median error over the whole detector. The blue line indicates an initial linear fit of the error over each slitlet. The orange line represents the model created in the first step of our wavelength calibration as described in Sect.~\ref{section:methods:wavelength} and shown in Fig.~\ref{fig:wavelength_calibration:third_order}. Panel (b): Slope of the linear fit (blue line in Fig.~\ref{fig:wavelength_calibration:across_slit}) to the wavelength error in each slitlet (black) as described in step 1 of Sect.~\ref{section:methods:wavelength}. The orange line shows the third-order polynomial fit to the data. Panel (c): Measurement of the wavelength error along the wavelength dimension within slitlet 11 (black) as described in step 2 of Sect.~\ref{section:methods:wavelength}. The orange line shows the spline fit to the data. The shaded areas correspond to the slitlets at the top and bottom of the field of view which are cropped out after the wavelength calibration. The figures at the top and on the bottom left illustrate step 1 whilst the figure on the bottom right shows step 2.}
\label{fig:wavelength_calibration}
\end{figure*}

Step 1: We first measure an initial shift in each spaxel of the data. This is done by cross-correlating each spaxel with the telluric template using the function \texttt{phase\_cross\_correlation} from \texttt{skimage} \citep{vanderWalt2014Scikit-image:Python} with an upsampling factor of 20. We found that the error varied approximately linearly over the extent of each slitlet (see the upper panel in Fig.~\ref{fig:wavelength_calibration}). Moreover, the slope of the error also varies from the slitlets on one side of the detector to the slitlets on the other side. This effect---which is hardly noticeable by eye on raw detector frames (cf. Fig.~\ref{fig:illustration_data})---is due to spectral curvature, which causes a line of constant wavelength to follow a curve from one side to the other. We modelled this spectral curvature by fitting the slope of the wavelength error over all slitlets using a third-order polynomial (see the lower left panel in Fig.~\ref{fig:wavelength_calibration}). We then integrated this third-order polynomial and fit the y-intercept of the resulting function to model the wavelength error within each slitlet (see the orange line in the upper panel in Fig.~\ref{fig:wavelength_calibration}). 
\looseness=-1

Step 2: We then measure higher-order wavelength errors. Since the telluric lines cover the whole extent of the K-long grating, we can locally measure the wavelength error by cross-correlating the telluric transmission template on subintervals of the spectrum. To do so, we first combine all spaxels within the same slitlet after correcting them by the initial error measured previously. We then define subintervals of size 200 wavelength channels that we shift by 50 channels along the wavelength dimension such that they overlap. The size of the intervals is chosen to ensure that enough telluric lines are contained for a robust measurement, while the overlaps increase the sampling along the wavelength dimension. We then cross-correlate each subinterval with the telluric transmission template. This yields a list of measurements of the wavelength error as a function of wavelength (see lower right panel in Fig.~\ref{fig:wavelength_calibration} for the calibration of the slitlet 11 taken as an example). We fit the error with a third-order spline and a smoothing factor of 0.4 using the function \texttt{UnivariateSpline} from \texttt{SciPy} \citep{Virtanen2020SciPyPython}. The smoothing factor was chosen by inspecting the resulting fits by eye and ensuring that noisy error measurements did not impact the spline. For the 100 channels at either side of the spectrum, we set the wavelength error as equal to the value of the spline at the last measurement point. This is done for the sake of robustness because the spline can grow polynomially on either side and can result in extreme values. 

Step 3: Finally, we update the wavelength solution. The standard pipeline provides an initial wavelength map computed from arc lamp images taken during daytime calibrations. This map relates each detector pixel to its corresponding wavelength. We update it by adding the errors from the two previous steps according to the column number and wavelength bin associated to each pixel. The updated wavelength map is then used together with the normal recipe \texttt{eris\_ifu\_jitter}.

\begin{figure*}[h!]
    \centering
    \includegraphics[width=0.95\hsize]{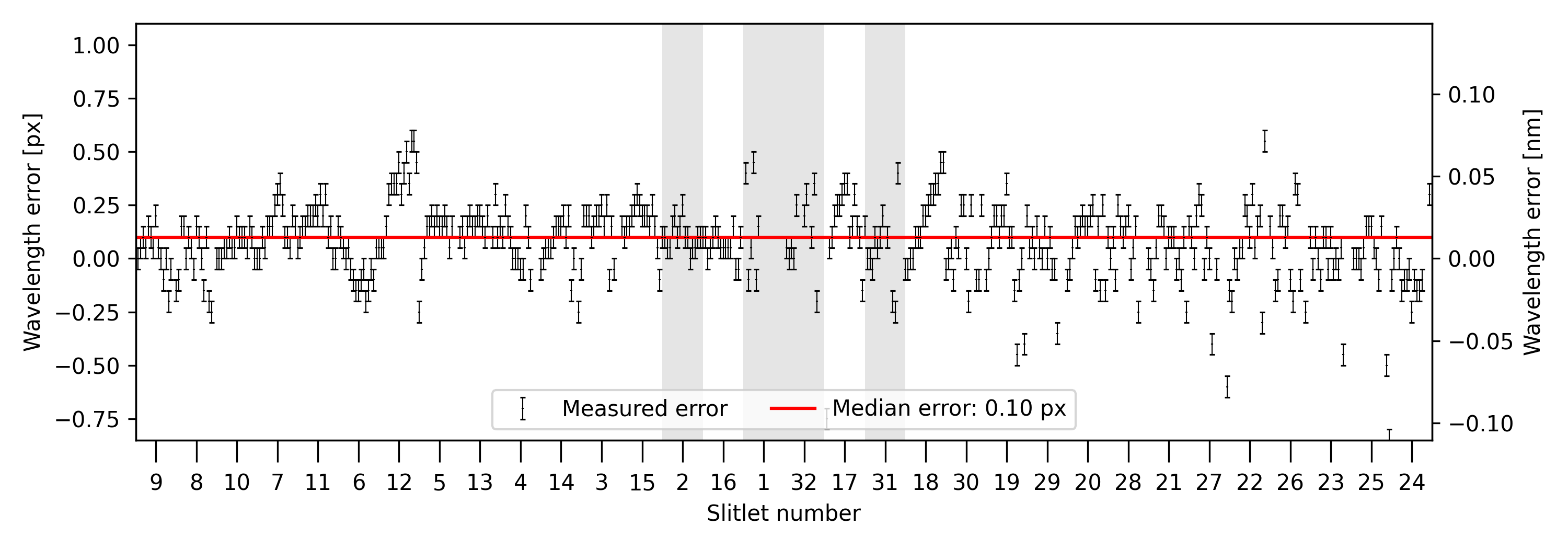}
    \caption{Validation of our custom wavelength calibration. The measurement described in step~1 of Sect.~\ref{section:methods:wavelength} was repeated after applying the wavelength calibration (black) to the same frame as in Fig.~\ref{fig:wavelength_calibration}. We only show every fourth measurement for readability. The shaded areas correspond to the slitlets at the edges of the field of view and are cropped out after the wavelength calibration.}
    \label{fig:wavelength_calibration_across_slit_control}
\end{figure*}

\begin{figure*}[ht]
\centering
    \begin{minipage}{0.49\textwidth}
        \centering
        \includegraphics[width=0.8\textwidth]{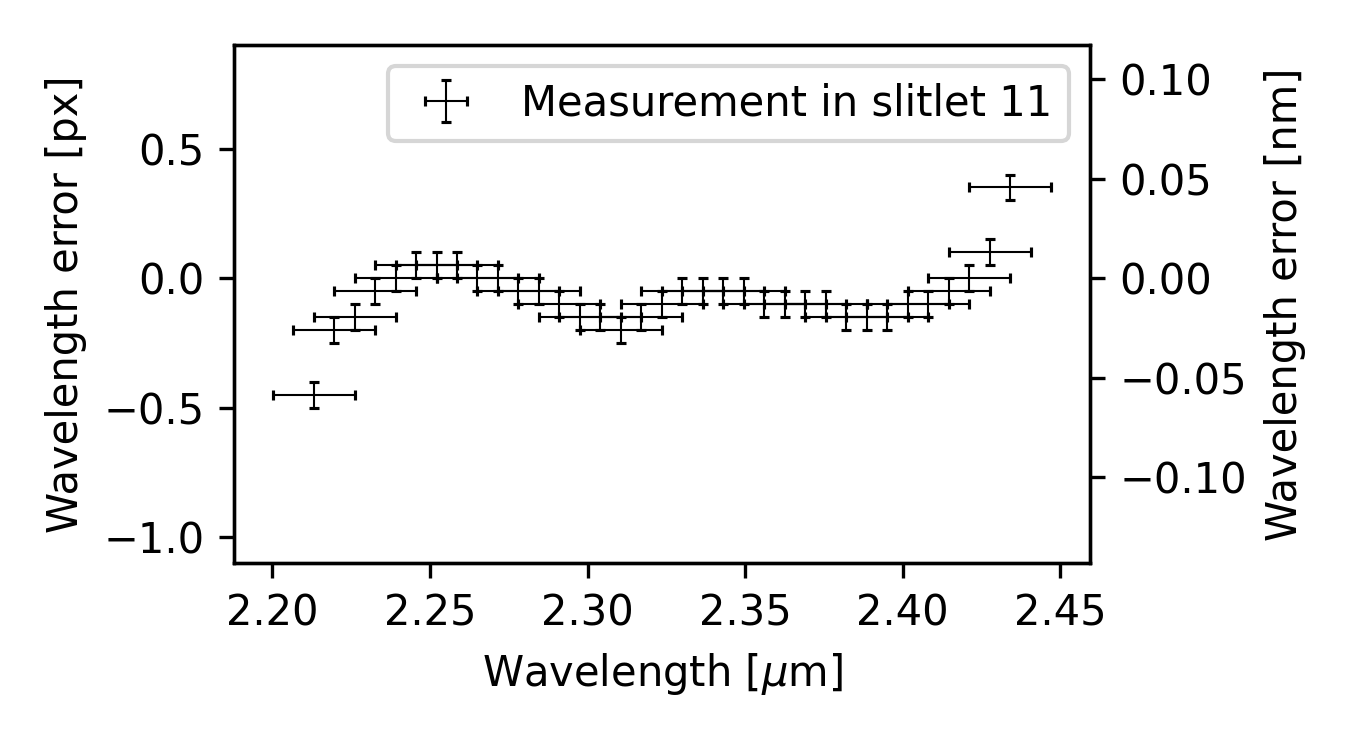}
        \caption{Validation of our custom wavelength calibration. The measurement described in step~2 of Sect.~\ref{section:methods:wavelength} was repeated after applying the wavelength calibration (black).}
        \label{fig:wavelength_calibration_spline_fit_control}
    \end{minipage}
    \hfill
    \begin{minipage}{0.49\textwidth}
        \centering
        \includegraphics[width=0.8\textwidth]{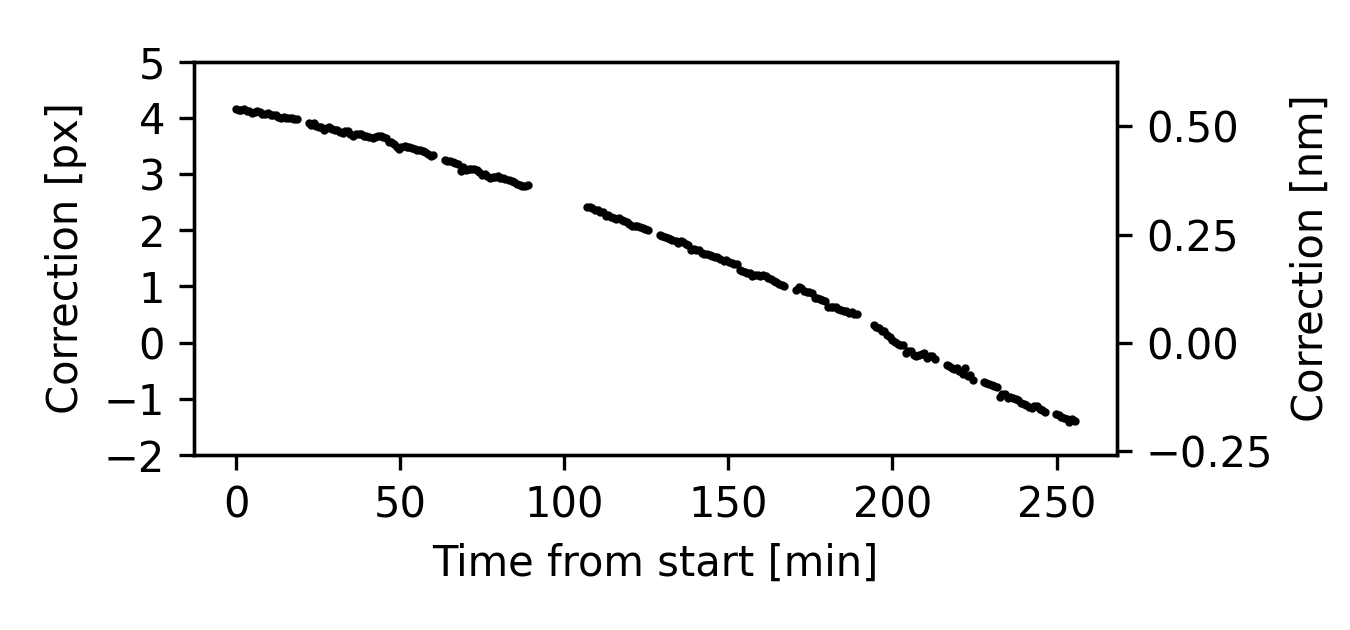}
        \caption{Median correction applied to the wavelength solution shown as a function of time since the start of the observation. The star was re-centred after 100~min, hence the break. The significant drift is due to flexures of the diffraction gratings of ERIS/SPIFFIER with changes in altitude and rotation.}
        \label{fig:wavelength_calibration_over_time}
    \end{minipage}
    
\end{figure*}

To validate our calibration, we repeat the measurement of the wavelength error described in steps 1 and 2 above after applying our custom calibration. Figures~\ref{fig:wavelength_calibration_across_slit_control} and \ref{fig:wavelength_calibration_spline_fit_control} present the results for the same frame as in Fig.~\ref{fig:wavelength_calibration}: most of the structure present in the uncalibrated data (cf. Fig.~\ref{fig:wavelength_calibration}) has been corrected, including both the linear trend across the slitlets and the wavy pattern along the wavelength axis. Unfortunately, some slitlets --- more precisely slitlets 12, 17, and 18---still show some trends of the order of 0.25 to \qty{0.5}{px}; however, the trend is significantly smaller than before the calibration. From a median error of \qty{3.85}{px} (i.e. \qty{0.5}{\nano\meter} or \qty{64}{\kilo\meter\per\second}) initially measured for the frame considered in Fig.~\ref{fig:wavelength_calibration}, our pipeline delivered a calibration with a median residual error of \qty{0.10}{px} (i.e. \qty{0.01}{\nano\meter} or \qty{1.7}{\kilo\meter\per\second}). 
To obtain an overview of the quality of the calibration, we repeated this exercise by measuring the median residual error of each frame after having applied our custom calibration. Over the whole observation, the median of the wavelength error is \qty{0.05}{px} and the standard deviation \qty{0.02}{px} (i.e. \qty{6.5+-2.4}{\pico\meter} or \qty{0.8+-0.3}{\kilo\meter\per\second}).
These results validate our custom wavelength calibration routine.
\looseness=-1

Over the whole observation, the wavelength solution varied by a total of \qty{5}{px}, i.e. \qty{0.65}{\nano\meter} or \qty{83}{\kilo\meter\per\second}, as shown in Fig.~\ref{fig:wavelength_calibration_over_time}. This large drift of the wavelength solution over time is due to internal flexures within SPIFFIER as the telescope---and thus the Cassegrain focus where ERIS is installed---changes altitude to track its target. The light-weighted diffraction gratings of SINFONI suffered from cryogenic deformation \citep{George2017ComplexGratings}, which significantly impacted its spectral profile. Therefore, heavier gratings were used for SPIFFIER which helped to greatly improve the spectral profile at the cost of significant flexures with changes in altitude and rotation. The impact of the flexures in SPIFFIER can be mitigated with a good wavelength calibration strategy---either by using OH sky emission lines \citep{Davies2007ASpectra} for faint targets and long integration times as is done by default by the ERIS-SPIFFIER pipeline, or by leveraging telluric absorption lines for observations of bright targets such as presented in this work.
\looseness=-1

%

\FloatBarrier

\section{Statistical test for the detection of molecules}
\label{app:detection_S/N}

\begin{figure*}[h]
    \centering
    \includegraphics[width=\hsize]{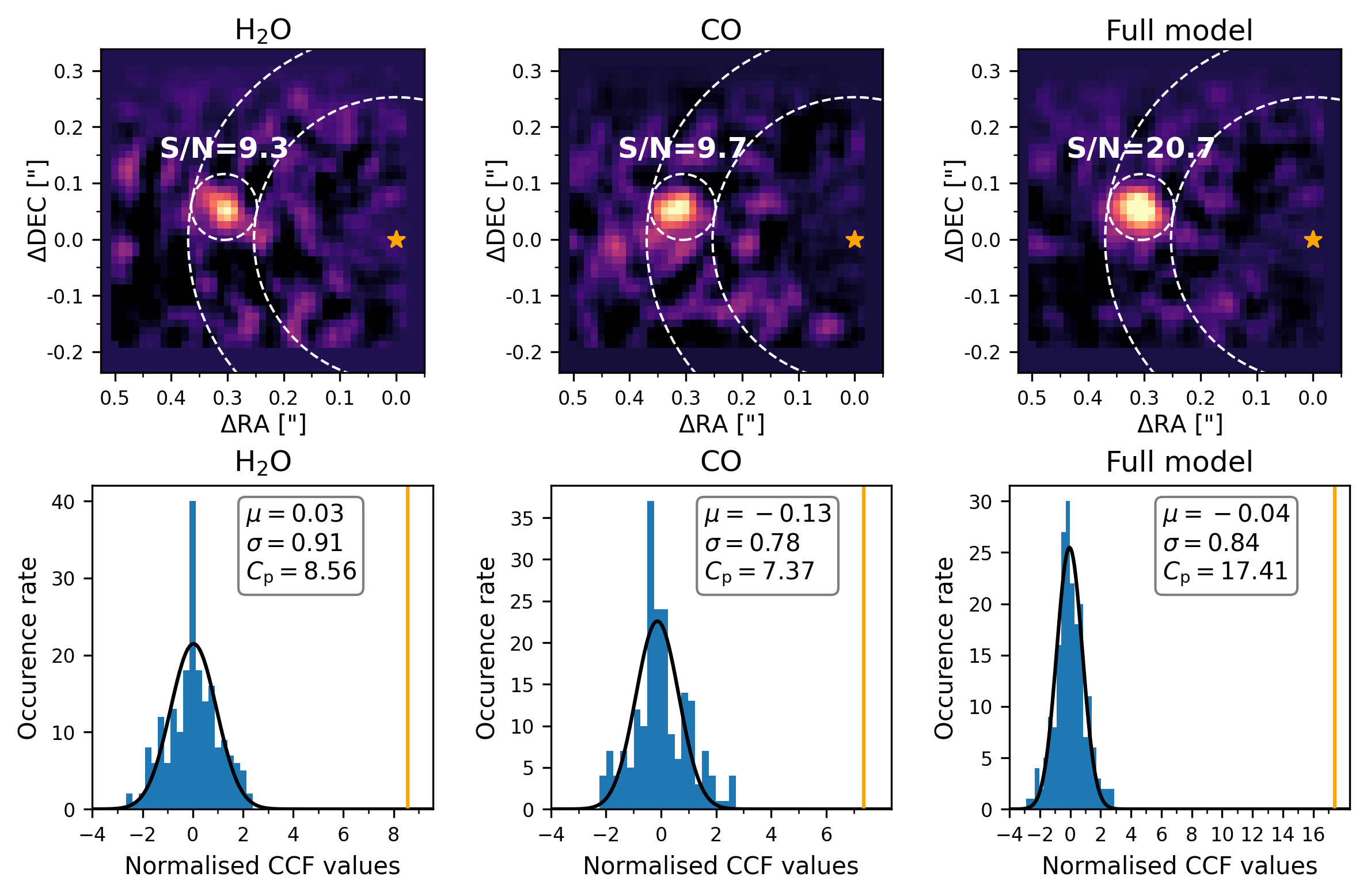}
    \caption{\textit{Top:} Molecular maps of H$_{2}$O, CO, and full model. The dashed annulus highlights the region considered to quantify the S/N of the detection. The dashed circle represents the position of AF~Lep~b and is masked for the purpose of quantifying the S/N. \textit{Bottom:} Histograms of the values of the CCF contained in the annular region. The dark line shows the Gaussian fit (with parameters $\mu$ and $\sigma$ shown in the legend), while the orange line shows the value $C_{\mathrm{p}}$ of the CCF at the position of the planet.}
    \label{fig:S/N_evaluation}
\end{figure*}

In this appendix, we describe the statistical test that we used to interpret the values of the cross-correlation function (CCF) as a measure for the confidence in the presence of molecular signal in each spaxel. This is done by setting a detection threshold on the values of the CCF based on a desired confidence level. Several such metrics have been used in the literature, as described in \citet{Garvin2024MachineNetworks}. 
We choose the test introduced by \citet{Petrus2021Medium-resolutionB}, as it has been used a few times already \citep{Patapis2022DirectMIRI,Malin2023SimulatedJWST/MIRI}, with a small modification to evaluate the CCF only at the angular separation of the planet. Concretely, we consider only the pixels of the standardised CCF (i.e. mean-subtracted and normalised by the standard deviation) at the RV of the planet (i.e. \qty{20}{\kilo\meter\per\second}) and in an annulus of width 1\,$\lambda/D$ (i.e. \qty{58.5}{\mas}) and radius equal to the angular separation of the planet (0\farcs30). We also mask out the 1~$\lambda/D$ region containing the planet. This results in a total of \qty{214}{px} considered for the noise estimate. We fit a Gaussian function to the histogram of their values. The S/N is then given by $\text{S/N} = (C_{\mathrm{p}}-\mu) / \sigma$, where $\mu$ and $\sigma$ are the mean and standard deviation of the fitted Gaussian, and $C_{\mathrm{p}}$ is the value of the CCF at the position of the planet. Figure~\ref{fig:S/N_evaluation} illustrates our measurement for the maps of H$_{2}$O, CO, and for the full model.

\FloatBarrier

\section{Measurement of the radial velocity}
\label{app:RV}

In this appendix, we describe the measurement of the RV of the planet. 

\textit{Preparation of the data:} We extracted two sets of spectra. We used an aperture of radius \qty{1.7}{px} (i.e. \qty{21.25}{\mas}) centred on the planet in the data both before and after PSF subtraction. The reason for extracting the stellar spectrum at the position of the planet before PSF subtraction is to investigate remaining calibration errors of the wavelength solution. We removed the continuum of the stellar spectrum using a Gaussian filter of size \num{40} (i.e. \qty{5.2}{\nano\meter}). 

\begin{figure*}[ht]
    \centering
    \includegraphics[width=0.85\hsize]{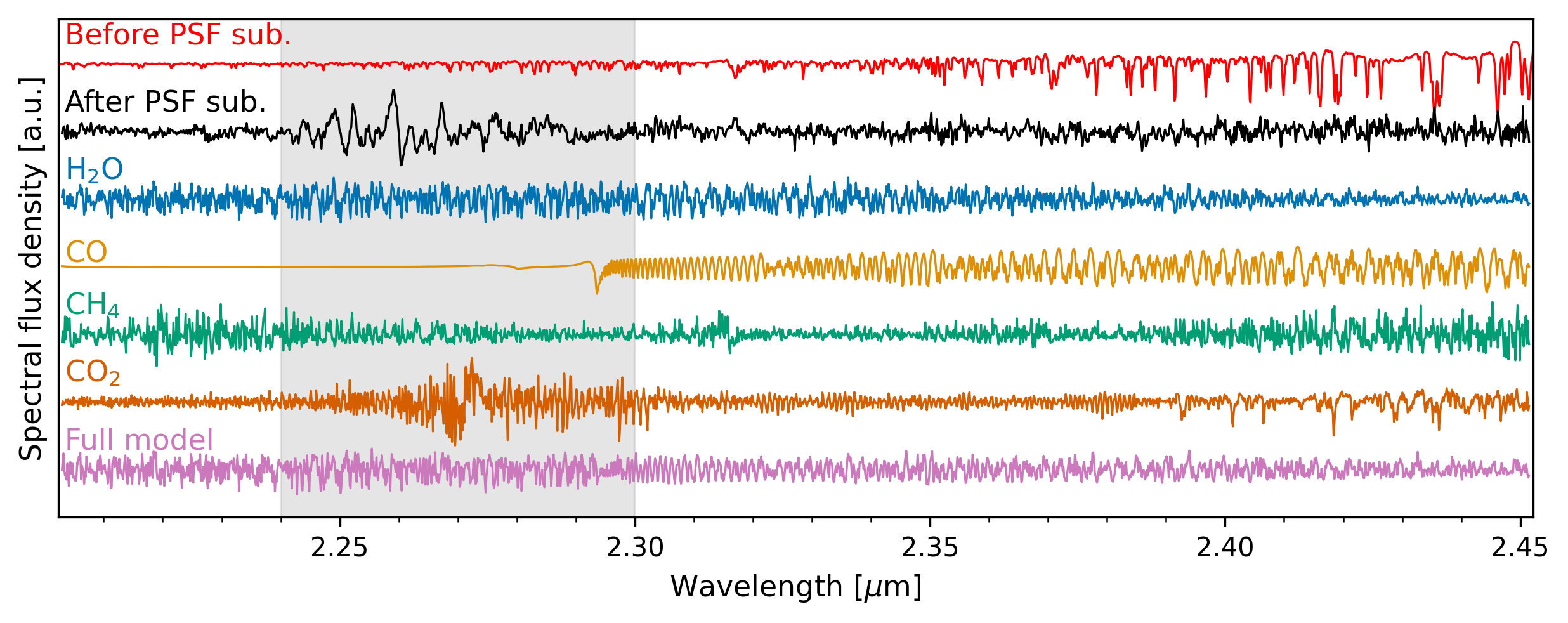}
    \caption{Spaxel at the position of the planet before and after PSF subtraction with spectral PCA and aperture convolution (\textit{red} and \textit{black}), and spectral templates used to compute the molecular maps presented in Fig.~\ref{fig:molmaps}. All spectra are shown after continuum subtraction. Additionally, they are normalised and offset vertically for readability. The shaded area between \qtyrange[range-phrase=~and~]{2.24}{2.3}{\micro\meter} represents the masked wavelength range where stray light is present.}
    \label{fig:spectrum}
\end{figure*}

\textit{Preparation of the models:} As a template for the telluric, we calculated a transmission spectrum of the Earth atmosphere with the same weather conditions as in the middle of our observations using SkyCalc-iPy,\footnote{\url{https://skycalc-ipy.readthedocs.io/en/stable/}} a Python wrapper to access ESO's SkyCalc tool \citep{Noll2012AnParanal,Jones2013AnParanal}. To obtain precise measurements of the RV, we used spectral templates at a higher bin sampling rate than the data (we chose a factor of 10) and convolved them down to the spectral resolution of the instrument using a Gaussian kernel with size $\sigma = (2\sqrt{2 \ln 2})^{-1}(\lambda / R) = \qty{0.055}{\nano\meter}$, where $\lambda = \qty{2.327}{\micro\meter}$ is the average wavelength of the K-long grating and $R = \num{11000}$ is the spectral resolution of ERIS/SPIFFIER. The continuum of the planet model was then removed using a Gaussian filter with the same physical size as the one used for the data, i.e. \qty{5.2}{\nano\meter}. We calculated the CCFs separately for each frame on the range \num{-80} to \qty{80}{\kilo\meter\per\second} and in steps of \qty{0.1}{\kilo\meter\per\second}. The spectral templates used to measure the RV of AF~Lep~b are shown in Fig.~\ref{fig:spectrum} together with the extracted spectrum of the planet before and after PSF subtraction.

\textit{Propagation of the uncertainty:} As described in the main text (see Sect.~\ref{section:results:RV}), we adopted a bootstrapping approach \citep{BeranAnBootstrap} to estimate the uncertainty arising from the systematics affecting the CCF. We describe the exact process in the following. We started by extracting the \num{239} planet spectra after PSF subtraction (i.e. before frame combination). We then cross-correlated these spectra with the spectral templates of H$_{2}$O, CO, and the full model.
We then picked \num{10000} bootstrap samples for each spectral template, each consisting of a random choice with replacement of the
\num{239} CCFs. For each sample, we mean-combined the CCFs and extracted the RV as the peak of the combined CCF. We noticed that, after bootstrapping, the measured RVs of the planet and the tellurics depended on the size of the filter and the method (Gaussian or median) used to remove the continuum. To remove this source of bias, we repeated the bootstrapping framework for both filter types and with filter sizes between \num{10} and \num{100} in steps of \num{2} (i.e. \qty{1.3}{\nano\meter} and \qty{13}{\nano\meter} in steps of \qty{0.26}{\nano\meter}), and pooled the resulting \num{112} sets of samples together, essentially weighting each filter equally.

\textit{Intermediate results:} We obtained the following values for the RV of the planet and for the spectral templates of H$_{2}$O, CO, and the full model: $v_{\mathrm{R,P}}^{(\mathrm{H_{2}O})} = \qty{20.6+-2.4}{\kilo\meter\per\second}$, $v_{\mathrm{R,P}}^{(\mathrm{CO})} = \qty{17.8+-2.7}{\kilo\meter\per\second}$, $v_{\mathrm{R,P}}^{(\mathrm{Full})} = \qty{19.5+-1.6}{\kilo\meter\per\second}$ and for the RV of the tellurics $v_{\mathrm{R,Tell}} = \qty{5.0+-0.2}{\kilo\meter\per\second}$. At first glance, the fact that the tellurics are not at rest in the data seems to invalidate the extensive wavelength calibration discussed in Sect.~\ref{section:methods:wavelength}. We report a systematic wavelength error of the order of \qty{0.8}{\kilo\meter\per\second} over the whole observation, which is six times smaller than the measurement of the RV of the tellurics reported here. We can resolve this contradiction by comparing the position of the planet in the molecular maps (Fig.~\ref{fig:molmaps}) with the residual wavelength error presented in Fig.~\ref{fig:wavelength_calibration_across_slit_control}. The planet is unfortunately located within the slitlet 18, which still shows some trend in the wavelength error from one side to the other side of the slitlet. Along slitlet 18, the remaining wavelength error after calibration varies by \qty{0.5}{px}, i.e. \qty{8.5}{\kilo\meter\per\second}. The planet being on the side of the slitlet, this causes our wavelength calibration to systematically leave a residual error in that location in the image. Fortunately, we can easily correct for this effect by subtracting the RV of the tellurics measured at the position of the planet from the RV of the planet. Future works could explore how to more accurately model the spectral curvature to better stabilise the wavelength calibration across the field of view of SPIFFIER in the cases not covered by the standard pipeline, i.e. when the stellar PSF dominates the image.

\textit{Calibration of the radial velocity:} To obtain the RV of AF~Lep~b relative to its host star, we first subtracted the RV of the tellurics to express our measurement in the rest frame of the observatory. We then transformed to the rest frame of the barycentre of the Solar System by correcting for the rotation of the Earth, the motion of the Earth around the Earth-Moon barycentre, and the motion of the Earth-Moon barycentre around the Sun. To do so, we used the function \texttt{helcorr} from \texttt{PyAstronomy} \citep{pya} at the Julian date of the middle of our observation \qty{2460256.72924}{\day}. This yielded a correction of $v_{\mathrm{R,BC}} = \qty{-13.96}{\kilo\meter\per\second}$ (as measured towards AF~Lep). We note that the correction varied between \num{-14.18} and \qty{-13.70}{\kilo\meter\per\second} between the start and end of our observations. We did not track the mean Julian date within each sampled frame used for our bootstrapping procedure; however, we expect the mean Julian date within each sample to vary by less than \qty{3}{\percent} due to the large number of frames (239) from which the samples are selected randomly, thus producing an additional uncertainty of the order of \qty{0.015}{\kilo\meter\per\second} (i.e. two orders of magnitude below our measurement uncertainty). For the RV of AF~Lep, we adopted the value measured by GAIA: $v_{R,\star} = \qty{20.61+-0.51}{\kilo\meter\per\second}$ \citep{Brown2018iGaia/i2}. Finally, the RV of AF~Lep~b relative to its host star is given by $\Delta v_{\mathrm{R,P\star}}^{(i)} = v_{\mathrm{R,P}}^{(i)} - v_{\mathrm{R,Tell}} - v_{\mathrm{R,BC}} - v_{R,\star}$, where $i$ denotes any of the three spectral templates used (i.e. H$_{2}$O, CO, and the full model).

\FloatBarrier

\section{Selected sample of fake planet injection tests}
\label{app:fake_planets}

\begin{figure*}[ht]
\centering
    \includegraphics[width=0.78\linewidth]{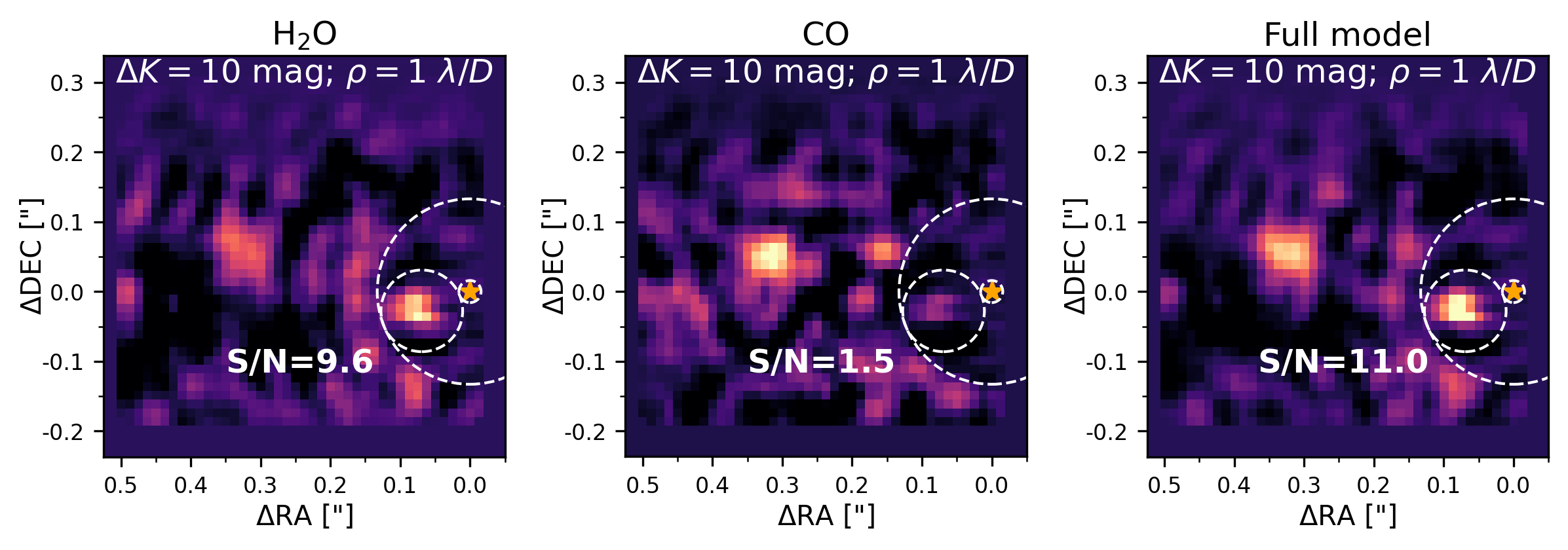}
    \includegraphics[width=0.78\linewidth]{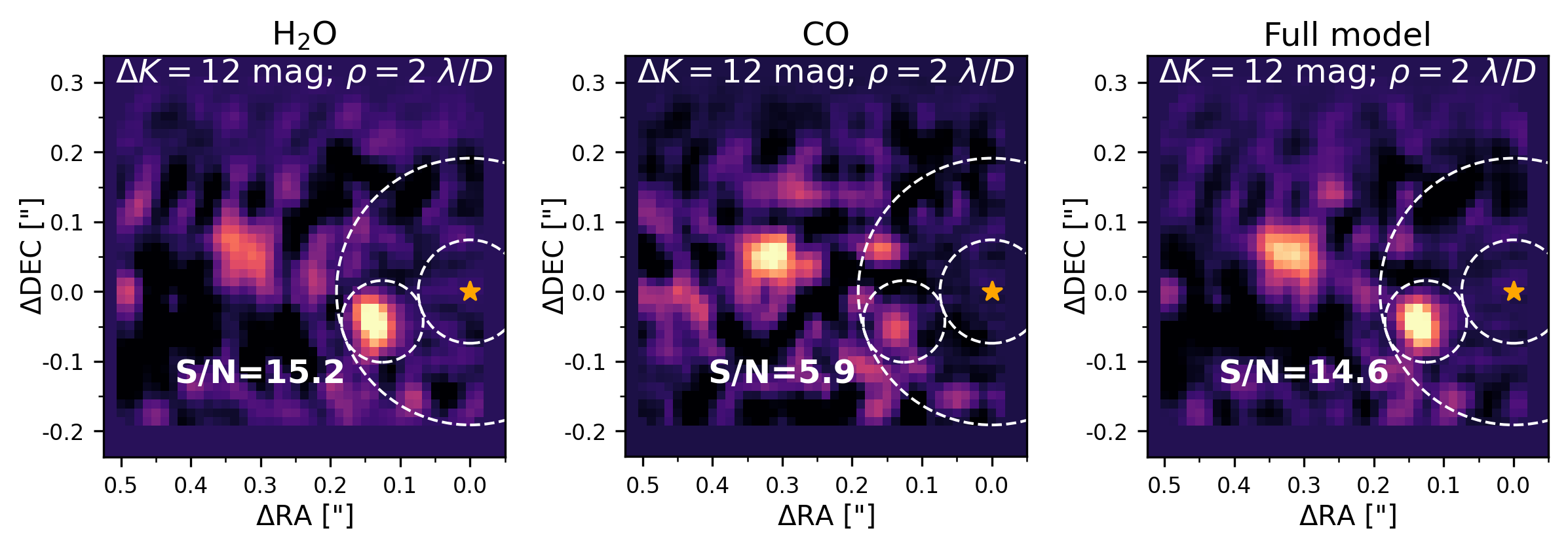}
    \includegraphics[width=0.78\linewidth]{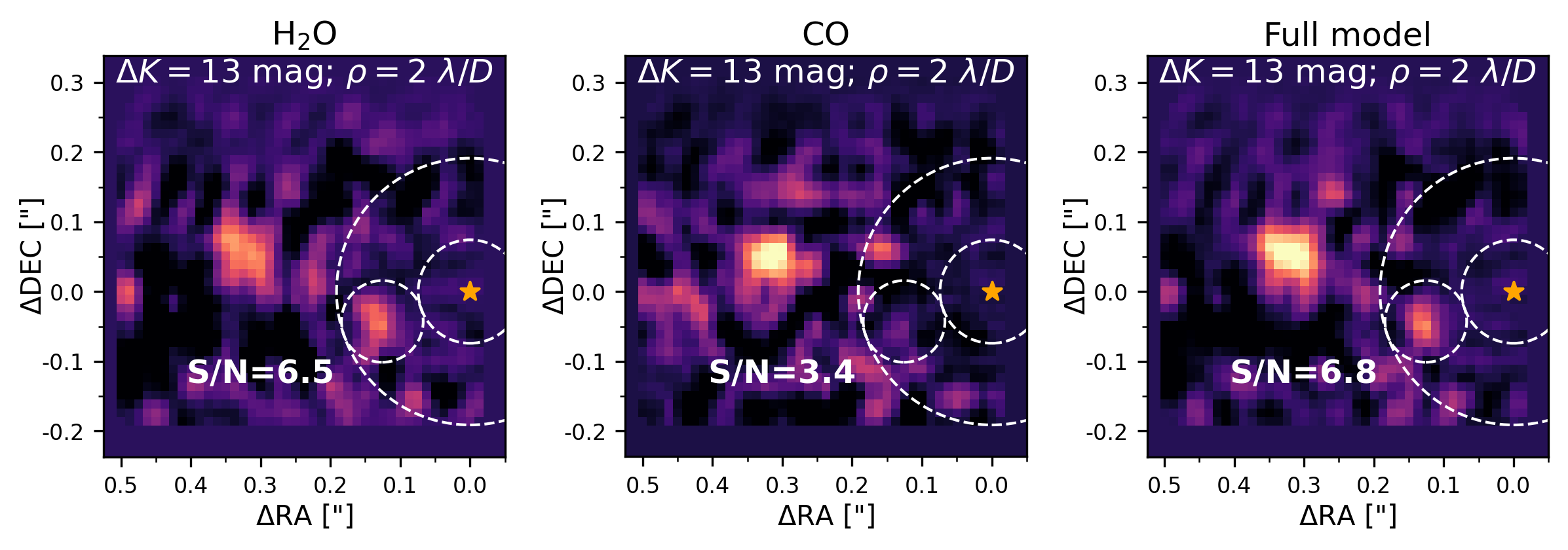}
    \includegraphics[width=0.78\linewidth]{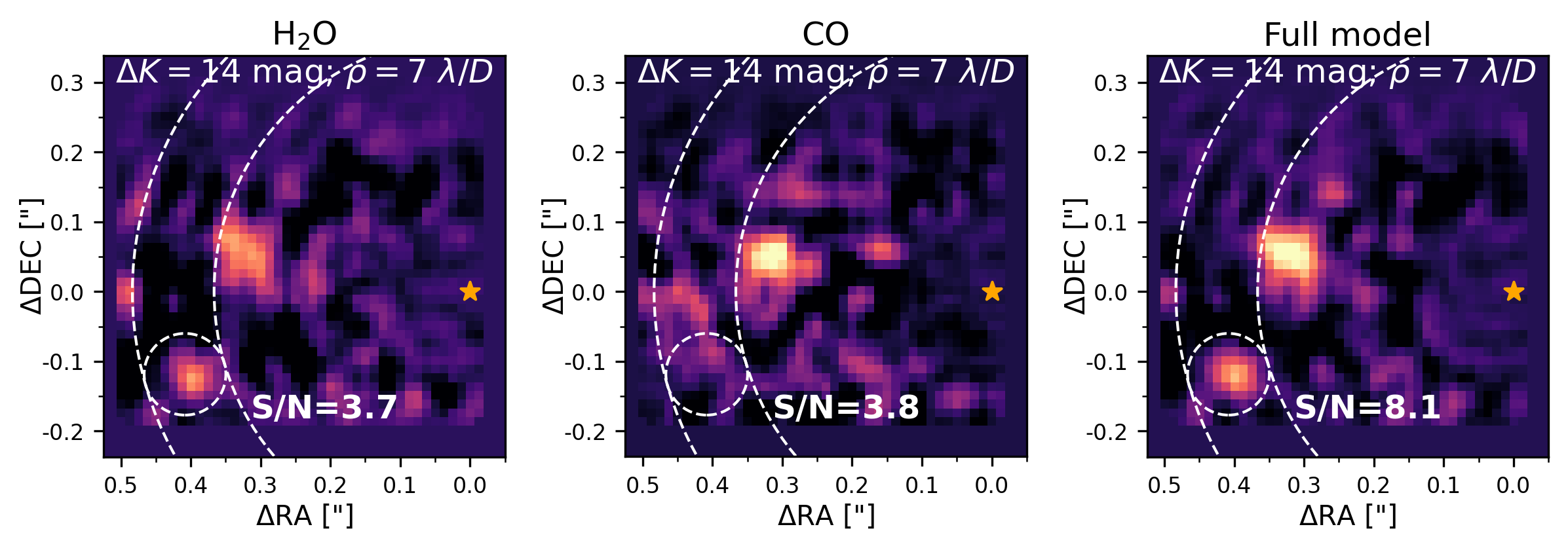}
\caption{Sample of fake planet injection tests to support the detection limits extracted from the contrast grids computed in Sect.~\ref{section:discussion:contrast_performance}. Each row shows a different experiment, and each column shows the corresponding molecular map computed with a different spectral template. The contrast and angular separation of the injected planets are annotated in white at the top of each figure. The dotted white circles show the position of the injected planets as well as the annulus considered to compute the S/N. The molecular maps show the cross-correlation values at the RV of the injected planet, i.e. \qty{0}{\kilo\meter\per\second}, which is why they look slightly different than Fig.~\ref{fig:molmaps}. The actual planet, AF~Lep~b, is still clearly visible due to the broad peak of the CCF (cf. Fig.~\ref{fig:ccf}).}
\label{fig:fake_planets}
\end{figure*}

\end{appendix}

\end{document}